\documentclass[%
 aip,
 amsmath,amssymb,
 reprint,%
]{revtex4-1}

\usepackage{graphicx}
\usepackage{dcolumn}
\usepackage{bm}

\usepackage[utf8]{inputenc}
\usepackage[T1]{fontenc}
\usepackage{mathptmx}
\usepackage{etoolbox}

\usepackage{comment}
\usepackage[mathlines]{lineno}
\usepackage{xcolor}
\usepackage{multirow}

\makeatletter
\def\@email#1#2{%
 \endgroup
 \patchcmd{\titleblock@produce}
  {\frontmatter@RRAPformat}
  {\frontmatter@RRAPformat{\produce@RRAP{*#1\href{mailto:#2}{#2}}}\frontmatter@RRAPformat}
  {}{}
}%
\makeatother
\begin{document}

\newcommand{\ex}[0]{\varepsilon_z}
\newcommand{\vs}[0]{\upsilon}
\newcommand{\fk}[0]{f_\kappa}
\newcommand{\xil}[0]{\xi_\lambda}

\preprint{}

\title[Ion temperature and lunar photoelectron sheath]{Effect of ion temperature on lunar photoelectron sheath}
\author{Trinesh Sana}
    \thanks{\url{sanatrinesh@gmail.com}.}
\affiliation{ 
Planetary Sciences Division, Physical Research Laboratory, Ahmedabad, 380009, India.
}%
\author{Sanjay K. Mishra}%
  \thanks{\url{nishfeb@gmail.com}.}
 \affiliation{ 
Planetary Sciences Division, Physical Research Laboratory, Ahmedabad, 380009, India.
}%

\date{\today}

\begin{abstract}
Observations suggest that the solar wind ions impinging on the lunar surface possess a finite temperature. The effect of ion temperature on the plasma sheath over the lunar surface is investigated in this paper. To account for thermal effects, a thermal pressure term has been added to the ion-flow force equation. Quantitative estimation of sheath characteristics has been performed by solving the Poisson equation, accounting for photoelectrons, solar wind electrons, and warm ions. Notably, the presence of warm ions within the sheath, i) changes the potential, field, sheath population density structures, ii) changes the sheath thickness, iii) reduces the photoelectron trapping, and iv) reduces solar wind electron reflection, compared to cold ions. The effect is more prominent at high lunar latitudes, which in turn may significantly modulate the dust dynamics. 
\end{abstract}

\maketitle
%
%
%
%
\section{Background and problem statement}
The characteristics of the lunar photoelectron sheath significantly depend on the distribution of photoelectrons, solar wind (SW)/ ambient plasma electrons, and ions above the charged lunar surface. As the UV-induced photoemission and SW electron collection are the dominant charging mechanisms on the lunar surface, a significant amount of research has been carried out considering different velocity distribution function (VDF)s of lunar photoelectrons and SW electrons \citep{grard1971photoelectron,feuerbacher1972photoemission,willis1973photoemission,manka1973plasma,walbridge1973lunar,nitter1998,poppe2010simulations,poppe2011modeling,sodhaandmishra2014,burinskaya2014influence,burinskaya2014influence,dyadechkin2015new,li2016dust,mishra2019photoelectron,mishra2020,zhao_photoelectron_2021,sana2023plasma}. However, in most of the previous analytical attempts, ions are treated cold within the sheath\citep{nitter1998,poppe2011modeling,burinskaya2014influence,burinskaya2015non,li2016dust,zhao_photoelectron_2021,sana2023plasma}. Due to the large supersonic drift speed of SW ions relative to their thermal speed, the cold-ion approximation is generally assumed good approximation, significantly simplifying the mathematical calculations. However, at higher lunar latitudes and the terminator region, the effective speed of SW ions normal to the surface is reduced to subsonic speed. Here, ions are accelerated to the Bohm speed $\sim\sqrt{k_BT_e/m_i}$ to maintain the sheath structure\citep{bohm1949}. $k_B$ is the Boltzmann constant, $T_e$ is the SW electron temperature, and $m_i$ is the ion mass.

\begin{figure}
    \centering
    \includegraphics[width=1\linewidth]{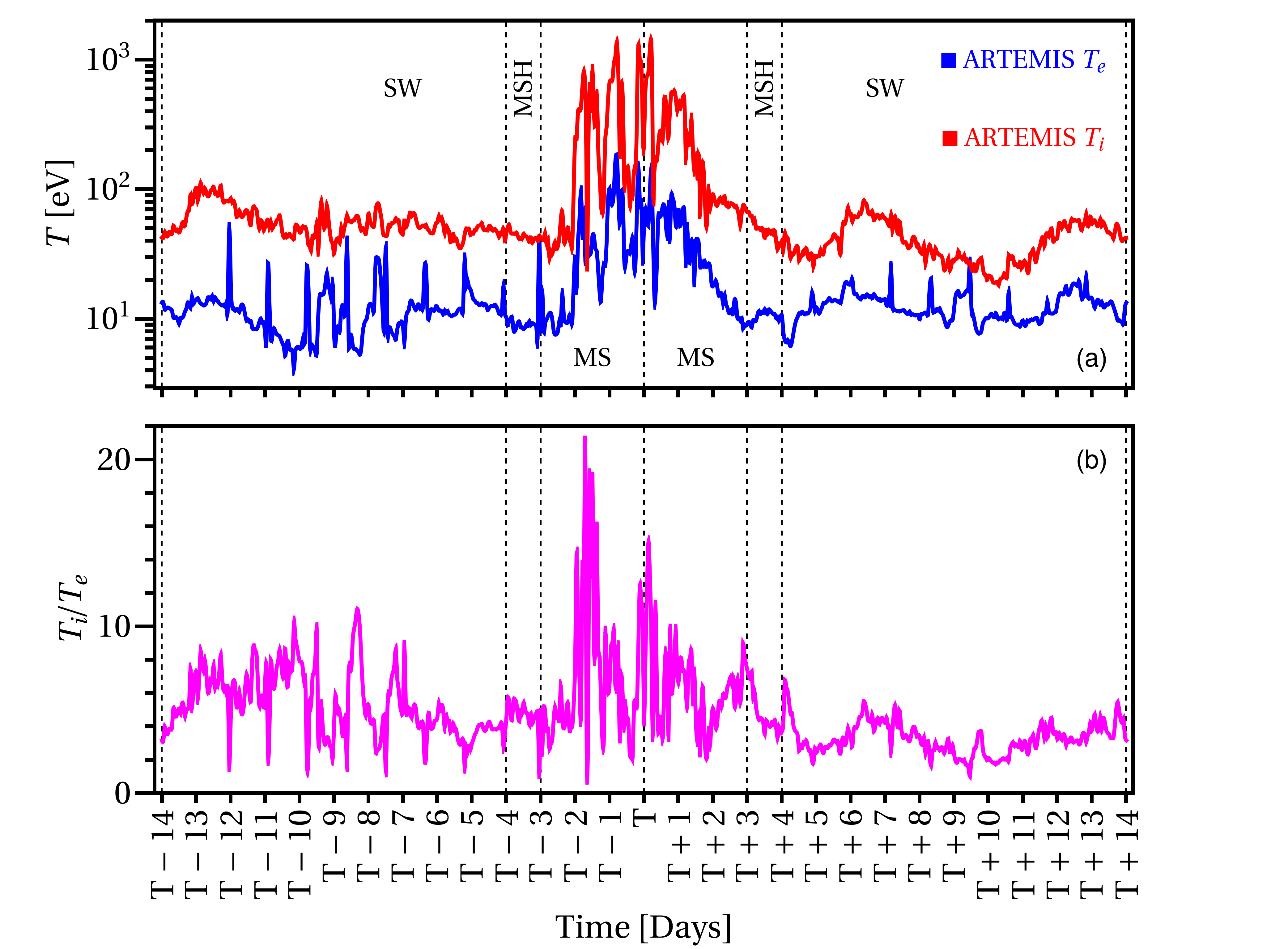}
    \caption{(a) Sample electron and ion temperature data measured by the THEMIS-C probe \citep{angelopoulos_artemis_2011}. Here, T is the time of the full Moon (2013-01-27 at 04:30 UTC). The vertical lines separate different regions: SW, solar wind; MSH, magnetosheath; MS: inner magnetosphere. (b) The variation of $T_i/T_e$ during Moon's passage through various plasma environments.}
    \label{fig:ARTEMIS}
\end{figure}
SW ions possess finite thermal speed observed by various missions such as WIND, THEMIS-ARTEMIS, ADITYA-L1 etc\citep{ogilvie1995swe,angelopoulos_artemis_2011,kumar2025aditya}. Figure \ref{fig:ARTEMIS} illustrates that during the passage through SW and Earth's magnetospheric plasma, 
Moon encounters ions with finite temperature, and $T_i/T_e$ varies in a wide range measured by the THEMIS-C probe\citep{angelopoulos_artemis_2011}. Hence, the cold-ion approximation does not account for this effect. The earlier particle-in-cell calculations of the lunar photoelectron sheath briefly considered the effect of the ion temperature\citep{poppe2010simulations,lisin2014effect,lisin2015lunar}. However, Ref.\cite{poppe2010simulations} considered SW with drift speed $u\sim400$ km s$^{-1}$ and equal electron and ion temperature, i.e., $(T_e=T_i)$. Although Refs.\cite{lisin2014effect,lisin2015lunar} considered $T_e\neq T_i$, they have investigated the dynamics  of supersonic ions $(u>>\sqrt{k_BT_e/m_i}\sim\sqrt{k_BT_i/m_i})$. Here, the effect of $T_i$ is marginal, and the cold-ion approximation is reasonably applicable. A recent work of Ref. \cite{basnet2025generalized} includes the effect of $T_i$ in a lunar dusty plasma scenario; however, they did not investigate the effect of $T_i$ in detail, as the major focus of their work was to address the influence of dust charge.

Overall, there is a notable lack in addressing the effect of $T_i$ on the lunar photoelectron sheath. Note that warm ions within the sheath modify the Bohm criterion\citep{bohm1949}. Hence, finite $T_i$ significantly affects the ion flux to the lunar surface and population density within the sheath. The analysis presented in this paper addresses the effect of finite ion temperature within the lunar photoelectron sheath. Over the sunlit region of the Moon, the present analysis provides a quantitative comparison of the sheath characteristics with warm and cold SW ions in terms of altitude profiles of the inherent electric potential, electric field, and population density.
%
%
\section{Lunar photoelectron sheath}
The lunar photoelectron sheath is a composite system of photoelectrons, solar wind (SW)/ ambient plasma electrons, and ions above the charged lunar surface. This is developed under the dynamic interaction of solar UV/EUV radiation and SW/ambient plasma with the lunar surface.  UV/EUV-induced photoemission and plasma collection lead to electrostatic charging of the lunar surface, resulting in a finite lunar surface potential\citep{manka1973plasma}. The effect of this surface potential is screened by photoelectrons and by the SW/ambient plasma, creating a non-neutral space-charge region above the sunlit lunar surface, forming the lunar photoelectron sheath\citep{sodhaandmishra2014}. The sheath exhibits inherent electric potential and field structures. The degree of surface charging and subsequent screening varies across different locations on the Moon, leading to distinct potential/field structures within the sheath. The formation of the lunar photoelectron sheath is widely discussed in the literature \citep{singer1962photoelectric,singer1962electrostatic,grard1971photoelectron,Fu_1970,fu_surface_1971,manka1973plasma,walbridge1973lunar,nitter1998,stubbs2007lunar,wang2008modeling,poppe2010simulations,poppe2011modeling,popel2013dusty,popel2014distributions,burinskaya2014influence,burinskaya2015non,dyadechkin2015new,sodhaandmishra2014,li2016dust,mishra2019photoelectron,mishra2020,zhao_photoelectron_2021,sana2023plasma}. The characteristics of the sheath structure depend on the population densities of photoelectrons, SW/ambient plasma electrons, and ions, and can be characterized by Poisson's equation, 

\begin{equation}
    \frac{d^2V}{dx^2}=-\frac{e}{\varepsilon_0}\big[n_i-n_e-n_p\big]\,,\label{eq:Basic Poisson Eq 1}
\end{equation}
where $e$ is the electronic charge, $\varepsilon_0$ is the permittivity in free space, $V$ is the electric potential at altitude $x$. $n_i$, $n_e$, and $n_p$ are the population densities of SW ions, SW electrons, and photoelectrons within the sheath, respectively. Poisson's equations connect the electric potential with the charge density. One needs to solve this equation with the boundary conditions $V(x=0)=V_0$ and $V'(x\rightarrow\infty)=0$ to quantitatively estimate the characteristics of the plasma sheath. 
\color{black}

On the locations with negligible photoemission (near the terminator and higher latitudes), the surface exhibits a negative surface potential due to dominant SW electron collection. Positively charged ions screen the surface potential, and the potential monotonically increases to zero with altitude, forming a classical Debye-type potential structure.  In the region with significant photoemission, the surface potential generally takes positive values, and screening is predominantly provided by photoelectrons. Here, the potential monotonically decreases to zero. However, the sheath exhibits another potential structure, in which the surface potential decreases to a negative minimum $V_m$ at $x_m$ and then increases to zero, giving rise to a non-monotonic potential structure. The non-monotonic potential structure performs two key functions: i) it traps the outgoing photoelectrons below potential minima $(x<x_m)$ and ii) it reflects incoming SW electrons above $x_m$. As a result, the non-monotonic solution is a more favorable steady-state configuration within the lunar photoelectron sheath in the region with significant photoemission \citep{nitter1998,poppe2010simulations,poppe2011modeling,sana2023plasma}. The existence of the non-monotonic potential structure has also been detected by Lunar Prospector and THEMIS-ARTEMIS\citep{halekas2008,halekas2011first-non-mono,poppe2011negative,poppe2012comparison}. To derive the non-monotonic solution from the Poisson equation, an additional condition has to be imposed to maintain zero electric field at the potential minima. 

\color{black}
The quantitative estimation of $V$ at different $x$ depends on the mathematical form of $n$ of different populations. $n$ significantly depends on the VDF of the constituent particle. The effect of different VDFs and subsequent $n(x)$ for photoelectrons and SW electrons on the lunar photoelectron sheath has been widely addressed in the literature. Note that ions play predominant roles in the screening of Debye-type potential structure and non-monotonic potential structure for $x>x_m$. However, the effect of $n_i(x)$ with finite $T_i$ has been poorly addressed in the context of the lunar photoelectron sheath. If a finite $T_i$ is introduced, the ion density within the sheath can be written as (see Appendix \ref{sec:Appendix nsi} for the detailed derivation) 

\color{black}
\begin{equation}
    n_i(x)=n_{i\infty}\left[1-\frac{2k_BT_e}{m_iu_{0}^2}\left\{\frac{eV(x)}{k_BT_e}+\psi\frac{T_i}{T_e}\ln\left(\frac{n_i(x)}{n_{i\infty}}\right)\right\}\right]^{-1/2}\,,\label{eq:ni_within_the sheath}
\end{equation}
where $m_i$ is the ion mass, $n_{i\infty}$, and $u_0=u_{i\infty}\cos\theta$ are the density and drift speed normal to the surface at the sheath edge. $T_{e(i)}$ is the electron (ion) temperature. $\theta$ is the subsolar angle (latitude). $\theta=0^\circ$ is the equator and $\theta=90^\circ$ represents the terminator. $\psi$ is the adiabatic index. For isothermal flow $\psi=1$ and $\psi=3$ for one dimensional adiabatic flow\citep{riemann1991bohm}. Eq. \ref{eq:ni_within_the sheath} can be solved numerically to estimate $n_i(x)$. The detailed derivation of $n_i$ is given in the Appendix \ref{sec:Appendix nsi}. In order to sustain the sheath structure, ions must enter the sheath with a certain speed called the Bohm speed\citep{bohm1949}. In the case of classical Debye sheath ions with finite $T_i$ and Maxwellian electrons, the Bohm speed can be written as \citep{riemann1991bohm} 

\begin{equation}
    u_B=\sqrt{\frac{k_BT_e}{m_i}\left(1+\psi\frac{T_i}{T_e}\right)}\,.\label{eq:Modifed_Bohm_Speed}
\end{equation}
Putting $T_i=0$ in Eqs. \ref{eq:ni_within_the sheath} and \ref{eq:Modifed_Bohm_Speed} restored the cold ion approximation of $n_i$ and $u_B$ used earlier by Refs\citep{nitter_dynamics_1992,nitter1998,poppe2011modeling,li2016dust,sana2023plasma}. Note that in the case of the lunar photoelectron sheath,  $u_B$ significantly depends on the distribution of the different sheath constituents. This may be illustrated from the generalized Bohm sheath criterion\citep{riemann1991bohm}

\begin{equation}
    \left[\left(\frac{dn_i}{dV}\right)-\left(\frac{dn_e}{dV}\right)-\left(\frac{dn_p}{dV}\right)\right]_{V\rightarrow0}\le0\,,\label{eq:Gen_Bohm_Cri}
\end{equation}
or,

\begin{equation}
    u_0^2\ge\frac{\psi k_BT_i}{m_i}+\frac{(n_{i\infty}/m_i)}{\left[\left(\frac{dn_e}{dV}\right)+\left(\frac{dn_p}{dV}\right)\right]_{V\rightarrow0}}=u_B^2\,.\label{eq:Gen_Bohm_Speed}
\end{equation}
As $n_e$ and $n_p$ depend on the sheath potential structure, the above equation highlights that $u_B$ also depends on $V_0$ and $V_m$. 

Ions should enter the sheath at a speed equal to or greater than the Bohm speed, i.e., $u_{i\infty}\cos\theta\ge u_B$. Hence, ion flux within the sheath gets modified as,

\begin{equation}
   I_i =n_{i\infty}u_0= n_{i\infty}\max\left[u_B,u_{i\infty}\cos\theta\right]\,,\label{eq:ion flux}
\end{equation}
The main goal of this analysis is to assess the effect of finite ion temperature. For simplicity, we consider the Maxwellian distributions of the emitted photoelectrons and SW electrons. The detailed expressions of $n_p$, $n_e$, boundary conditions, computation schemes, and parameters used to solve Eq. \ref{eq:Basic Poisson Eq 1} are given in the Appendix \ref{sec:Appendix npe and nse}, \ref{sec:BC}, and \ref{sec:parameter}.
%
%
\section{Numerical results and discussions}
\subsection{Effect of $T_i$ on the classical Debye sheath}

\begin{figure}
    \centering
    \includegraphics[width=1\linewidth]{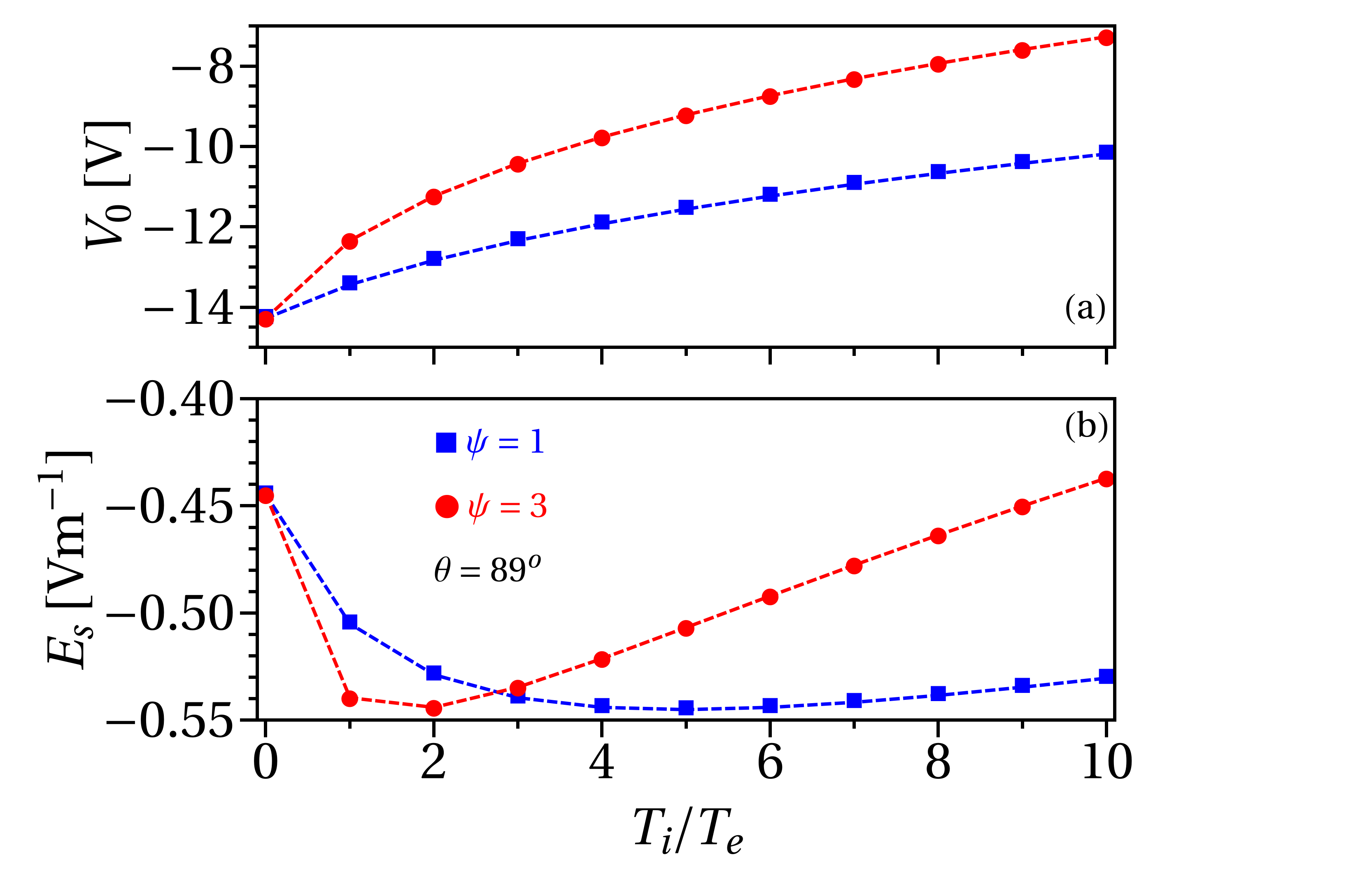}
    \caption{Variation of lunar surface potential $(V_0)$ and surface electric field $(E_s)$ with $T_i/T_e$ for classical Debye type potential structure at $\theta=89^\circ$. Other parameter used: $n_{i\infty}=5$ cm$^{-3}$, $T_e=15$ eV, $u_{i\infty}=400$ km s$^{-1}$, and $J_{ph0}=9.7\,\mu $A m$^{-2}$.}
    \label{fig:V0 and Es Type C at 89}
\end{figure}

\begin{figure}
    \centering
    \includegraphics[width=1\linewidth]{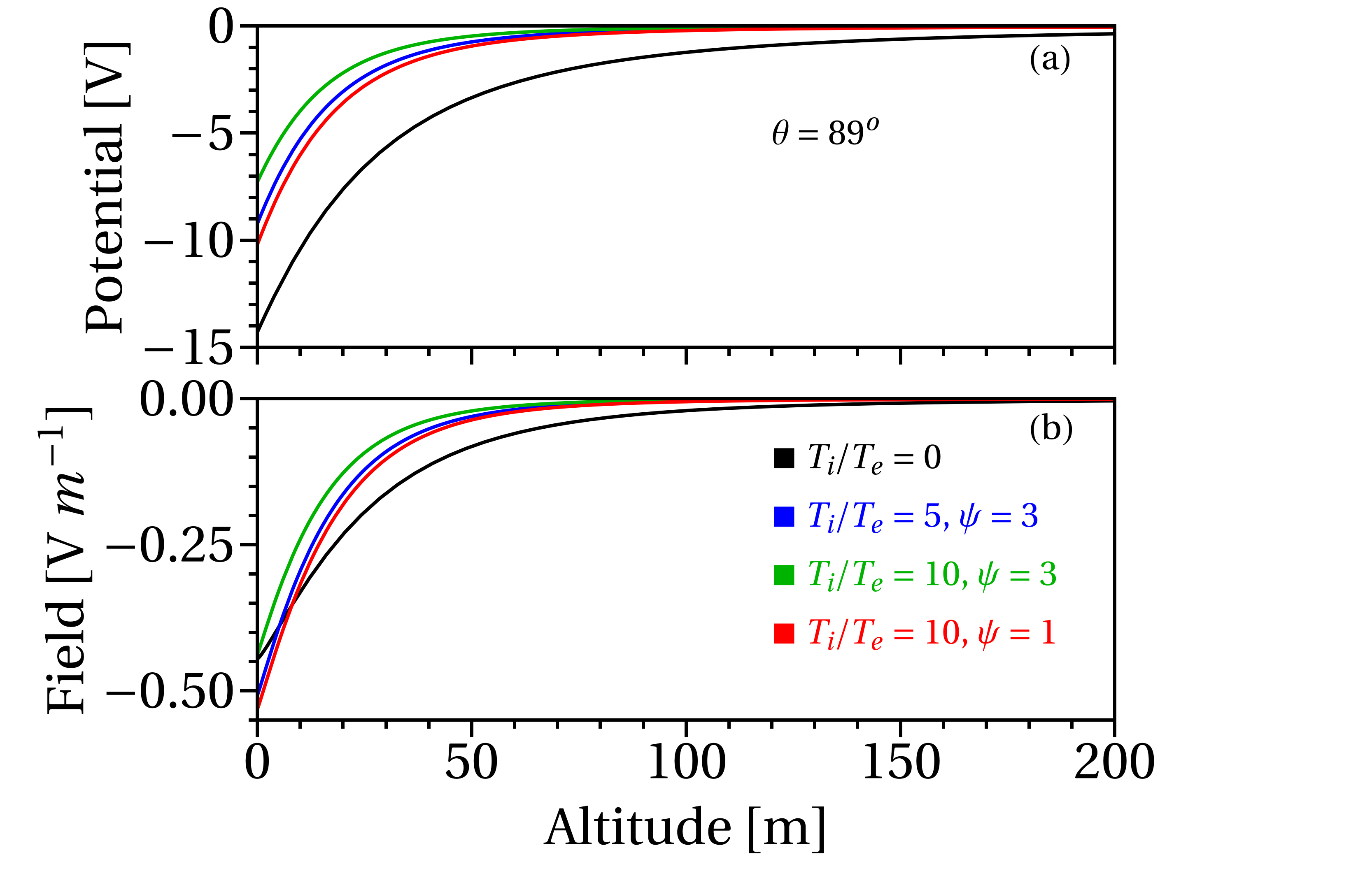}
    \caption{Altitude profile of electric potential and field within classical Debye type sheath structure at $\theta=89^\circ$ for different $T_i/T_e$ and $\psi$. Other parameter used: $n_{i\infty}=5$ cm$^{-3}$, $T_e=15$ eV, $u_{i\infty}=400$ km s$^{-1}$, and $J_{ph0}=9.7\,\mu $A m$^{-2}$.}
    \label{fig:Type C at 89}
\end{figure}
In the terminator region of the sunlit Moon, solar irradiation falls at a grazing angle, resulting in marginal photoemission, and $u_{i\infty}\cos\theta$ falls below $u_B$. Hence, lunar surface charging is primarily influenced by the collection of SW electrons, resulting in a negative surface potential. This negative potential is screened by the SW ions. Therefore, in these areas, $T_i$ is crucial for accurate estimation of potential structures. Figure \ref{fig:V0 and Es Type C at 89} illustrates the variation of lunar surface potential and surface electric field with $T_i/T_e$ for classical Debye-type sheath structure at $\theta=89^\circ$. The corresponding altitude profiles of potential and field are illustrated in Figure \ref{fig:Type C at 89}. Here, ions enter the sheath with $u_B$ (the sheath's natural response), which increases with increasing $T_i$ (see Eq. \ref{eq:Gen_Bohm_Speed}). This increases the effective ion flux on the lunar surface, leading to a decrease in the magnitude of negative potential with increasing $T_i$ and $\psi$ (as shown in Figure \ref{fig:V0 and Es Type C at 89}a and \ref{fig:Type C at 89}a). The lower negative potential (in magnitude) screens out at a lower altitude. 

\begin{figure}
    \centering
    \includegraphics[width=1\linewidth]{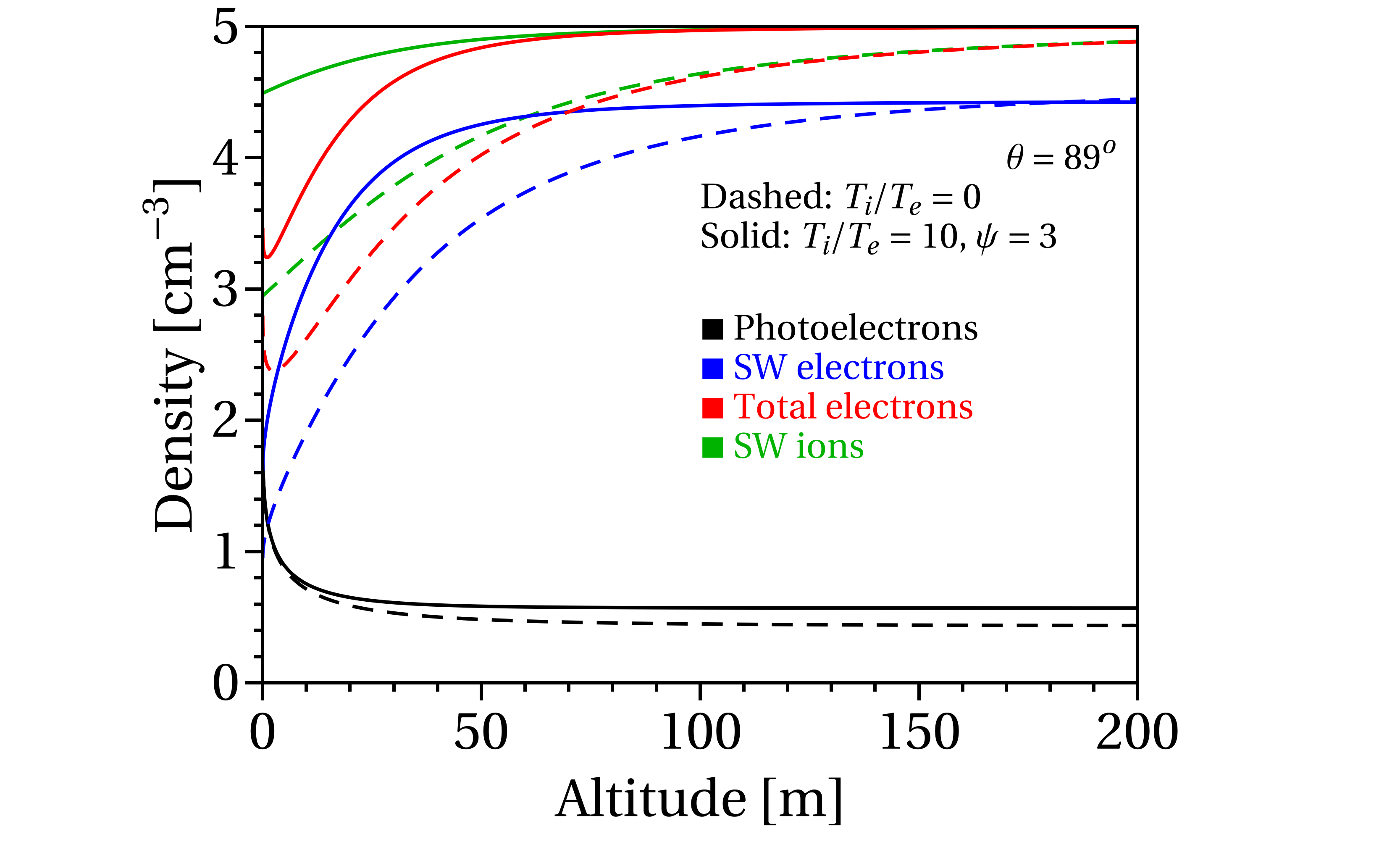}
    \caption{Altitude profile of population densities within classical Debye type sheath structure at $\theta=89^\circ$ for $T_i/T_e =0$ (Dashed) and $T_i/T_e =10$, $\psi=3$ (Solid). Other parameter used: $n_{i\infty}=5$ cm$^{-3}$, $T_e=15$ eV, $u_{i\infty}=400$ km s$^{-1}$, and $J_{ph0}=9.7\,\mu $A m$^{-2}$.}
    \label{fig:Density_Type C at 89}
\end{figure}
Furthermore, due to the thermal pressure, $n_i$ decreases less rapidly within the sheath compared to the cold ion case, as shown in Figure \ref{fig:Density_Type C at 89}. As a result, the sheath with warm ions exhibits a higher population of ions compared to cold ions. This results in a decrease in the magnitude of the negative lunar surface potential and lower screening altitude due to an increase in the effective ion flux and higher ion population within the sheath. As illustrated in Figure \ref{fig:V0 and Es Type C at 89}b, we observed that the magnitude of the negative surface electric field initially increases with increasing $T_i/T_e$ for fixed $\psi$. However, the surface electric field strength reduces for larger $T_i/T_e$. An initial increase of $T_i$ gives rise to a higher screening due to higher ion density within the sheath, which increases the field strength. However, for larger $T_i$ the magnitude of $V_0$ reduces and results in a weaker surface electric field. At a higher altitude, the overall negative electric field is observed to be weaker for the sheath with warm ions than with cold ions. Therefore, including $T_i$ in the estimates of the plasma environment surrounding the termination region is crucial, especially in the estimate of charged dust dynamics. Next, we will discuss the effect of $T_i$ in the region with significant photoemission.  
\subsection{Effect of $T_i$ on the non-monotonic sheath}
A non-monotonic sheath structure with a finite surface potential $V_0$ and negative potential minima $V_m\le0$ at finite altitude $x_m$ prevails around the region with significant photoemission. It traps most of the outgoing photoelectrons between $x=0$ and $x=x_m$, creating a potential barrier of $V_0-V_m>0$; and deflects incoming electrons from the SW/ ambient plasma for $x\ge x_m$ due to $V_m$. At nominal SW plasma conditions, for lower values of $\theta$, significant photoemission dominantly contributes to the estimation of the sheath structure. As a result, $V_0$ acquires positive values, and the magnitude of the negative potential minima remains small. At higher latitudes, both photoemission and SW electron collection have comparable contributions to the lunar surface charging. Here, $V_m$ acquires a value approximately of the order of $T_e$. And $V_0$ can take small positive or negative values, while keeping $V_0-V_m>0$ on the order of $T_p$ to trap most of the photoelectrons.

Ions also play a significant role in the formation of non-monotonic potential structure. For $x\ge x_m$, SW ions dominate over photoelectrons and SW electrons to screen the effect of potential minima $V_m$. To fulfill the Bohm criterion (as shown in Eq. \ref{eq:Gen_Bohm_Cri}), ions enter the sheath with sufficient speed so that $n_i$ falls more slowly than $n_e$ and $n_p$ near the sheath edge\citep{riemann1991bohm}. Note that Eq. \ref{eq:Gen_Bohm_Cri}, only contains the information near the sheath edge. However, for a non-montonic sheath, ions must overcome a repulsive potential barrier of $V_0-V_m$ to reach the lunar surface. We observed that for certain positive values of $V(x)$, ions entering the sheath with $u_B$ do not exhibit any real value of $n_i(x)$ near the lunar surface. To explain these observations in detail, let us first consider the case of cold ions. Putting $T_i=0$ in Eq. \ref{eq:ni_within_the sheath}, the number density of ions at any $x$ within the sheath can be written as

\begin{equation}
    n_i(x)=n_{i\infty}\left[1-\frac{2eV(x)}{m_iu_0^2}\right]^{-1/2}\,.
\end{equation}
The above equation gives the real values of $n_i(x)$ only for

\begin{equation}
    \frac{1}{2}m_iu_0^2>eV(x)\,,
\end{equation}
when $V(x)$ is positive. Otherwise, ions will not reach that layer. Putting $u_0=u_B=\sqrt{k_BT_e/m_i}$ for cold ions, we get

\begin{equation}
    V_c=\frac{k_BT_e}{2e}\,.\label{eq:Vc_Cold_Ion}
\end{equation}
If the positive potential $V(x)$ within the sheath is greater than $V_c$, $n_i(x)$ does not have a real value, which violates the ion continuity equation within the sheath. Hence, the sheath solution can not mathematically exist. For nominal SW conditions, with $T_e=15$ eV, $V_c$ is $+7.5$ V. This much positive potential develops near the equatorial region of the Moon due to dominant photoemission. However, in this region, SW ions have a supersonic speed $(\sim 400$ km s$^{-1})$, which makes $m_iu_0^2/2>>eV_c$. As a result, $n_i(x)$ has real roots everywhere within the sheath. At higher latitudes, although the SW speed falls to subsonic, due to a decrease in photoemission flux, if the surface gets a positive value, $V_0$ always remains below $V_c$. Hence, if ions are treated as cold, under nominal solar irradiation and SW/ambient plasma conditions, a non-monotonic sheath solution prevails around the Moon.

The situation becomes significantly different for ions with $T_i$. $V_c$ in the case of warm ions is smaller than that for cold ions (please see Appendix \ref{sec:Vc for warm ions} for details about $V_c$ for warm ions). Due to this low value of $V_c$, at higher latitudes, a small positive potential $V(x)>V_c$ leads to no real $n_i(x)$ values close to the surface if ions enter the sheath with $u_B$. In such cases, to retain the steady-state sheath structure, we propose to modify the Bohm sheath as  

\begin{equation}
    \tilde u_B^2=u_B^2+\eta^2\,.\label{eq:Bohm with push eta}
\end{equation}
$\eta$ is expected to be small compared to $u_B$. Warm ions entering the sheath with speed equal to or greater than $\tilde u_B$ can reach the lunar surface, and real values of $n_i(x)$ exist within the non-monotonic sheath structure.

\begin{figure}
    \centering
    \includegraphics[width=1\linewidth]{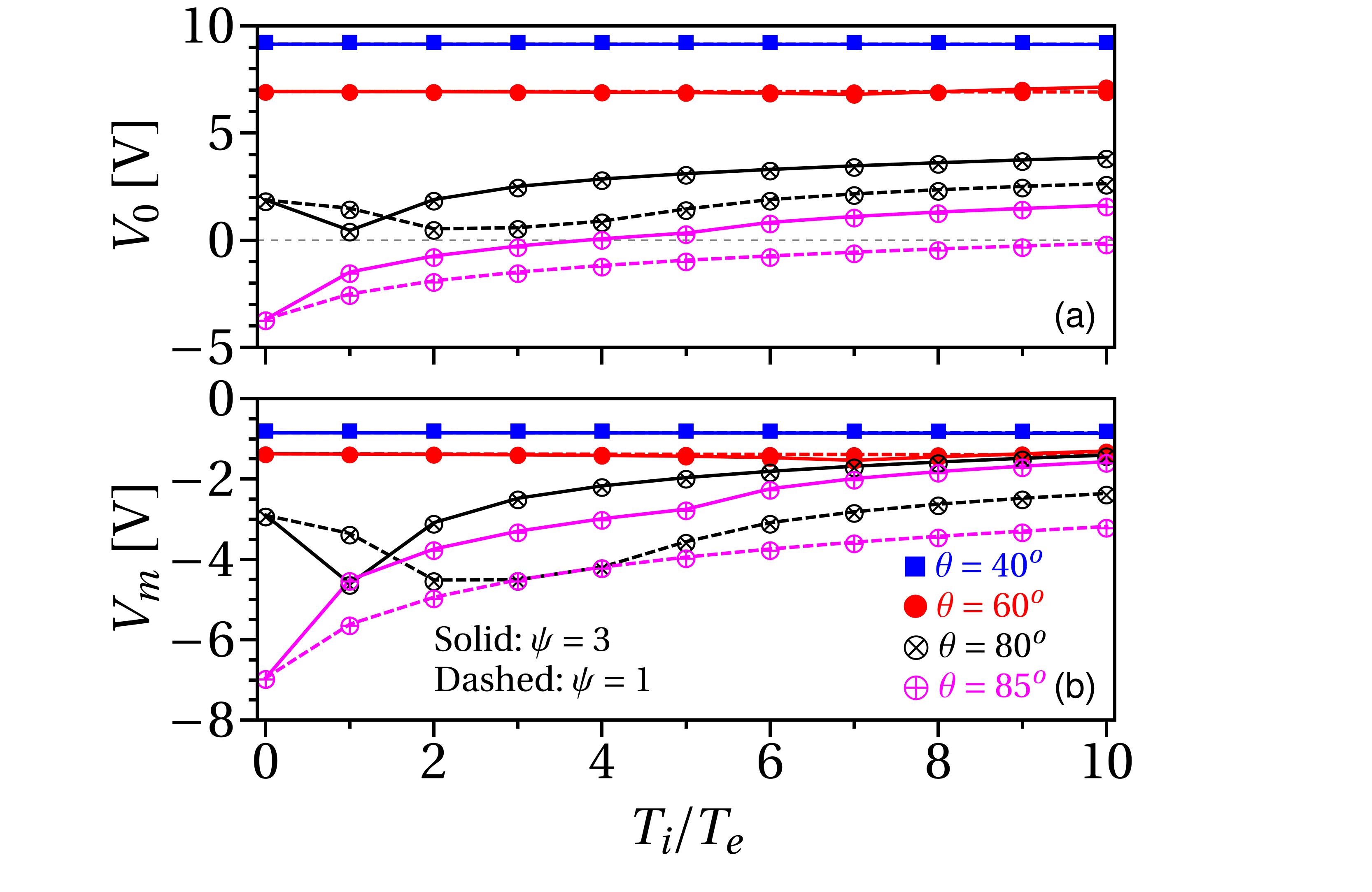}
    \caption{Variation of $V_0$ and $V_m$ with $T_i/T_e$ for non-monotonic potential structure for different $\theta$. Other parameter used: $n_{i\infty}=5$ cm$^{-3}$, $T_e=15$ eV, $u_{i\infty}=400$ km s$^{-1}$, and $J_{ph0}=9.7\,\mu $A m$^{-2}$.}
    \label{fig:V0 and Vm variation for Type A&B}
\end{figure}

Figure \ref{fig:V0 and Vm variation for Type A&B} illustrates the variation of lunar surface potential $(V_0)$ and potential minima $(V_m)$ for non-monotonic potential structures with $T_i/T_e$ within the lunar photoelectron sheath at different latitudes. At lower latitudes with smaller $\theta$ (blue curves in Figure \ref{fig:V0 and Vm variation for Type A&B}), no significant difference is observed in $V_0$ and $V_m$ for different $T_i/T_e$. This can be attributed to the fact that, at these $\theta$ values, photoemission is significantly high. Additionally, $u_{i\infty}\cos\theta$ is much greater than $\tilde u_B$, which results in a marginal contribution of $T_i$ in the lunar surface charging and subsequent sheath formation. A slight decrease in $V_0$ and $V_m$ with increasing $T_i/T_e$ is observed near $T_i/T_e=1$, $\psi=3$ for $\theta=80^\circ$ (the black solid curves in Figure \ref{fig:V0 and Vm variation for Type A&B}a \& b). Here, $u_{i\infty}\cos\theta$ remains greater than $\tilde u_B$ and ions enter the sheath with drift speed $u_{i\infty}\cos\theta$. Hence, $T_i$ does not affect the flux balance within the sheath. $T_i$ only contributes to the determination of potential minima $V_m$ through $n_i$. In such cases, a finite $T_i$, the thermal pressure gradient reduces $n_i$'s contribution to the determination of $V_m$. As a result, $V_m$ decreases and further decreases $V_0$. Similar thing is observed at $\theta=80^\circ$ for $T_i/T_e<3$, $\psi=1$. 

As $u_{i\infty}\cos\theta$ drops below $u_B$, $T_i$ actively participates both in the flux balance and in the determination of $V_m$. In such cases, ions are accelerated to $\tilde{u}_B$, and the thermal pressure gradient increases the contribution of $n_i$ in the determination of $V_m$. Therefore, the cumulative effect of $T_i$ increases the value of $V_0$ and $V_m$ (i.e., decreases the magnitude of $V_m$) with increasing $T_i/T_e$. This effect is prominently illustrated in magenta curves in Figure \ref{fig:V0 and Vm variation for Type A&B} for $\theta=85^\circ$. Overall, $T_i$ has a significant influence on the estimates of lunar surface potential $V_0$ and potential minima $V_m$ corresponding to the non-monotonic potential structure, especially at higher latitudes.

\begin{figure}
    \centering
    \includegraphics[width=1\linewidth]{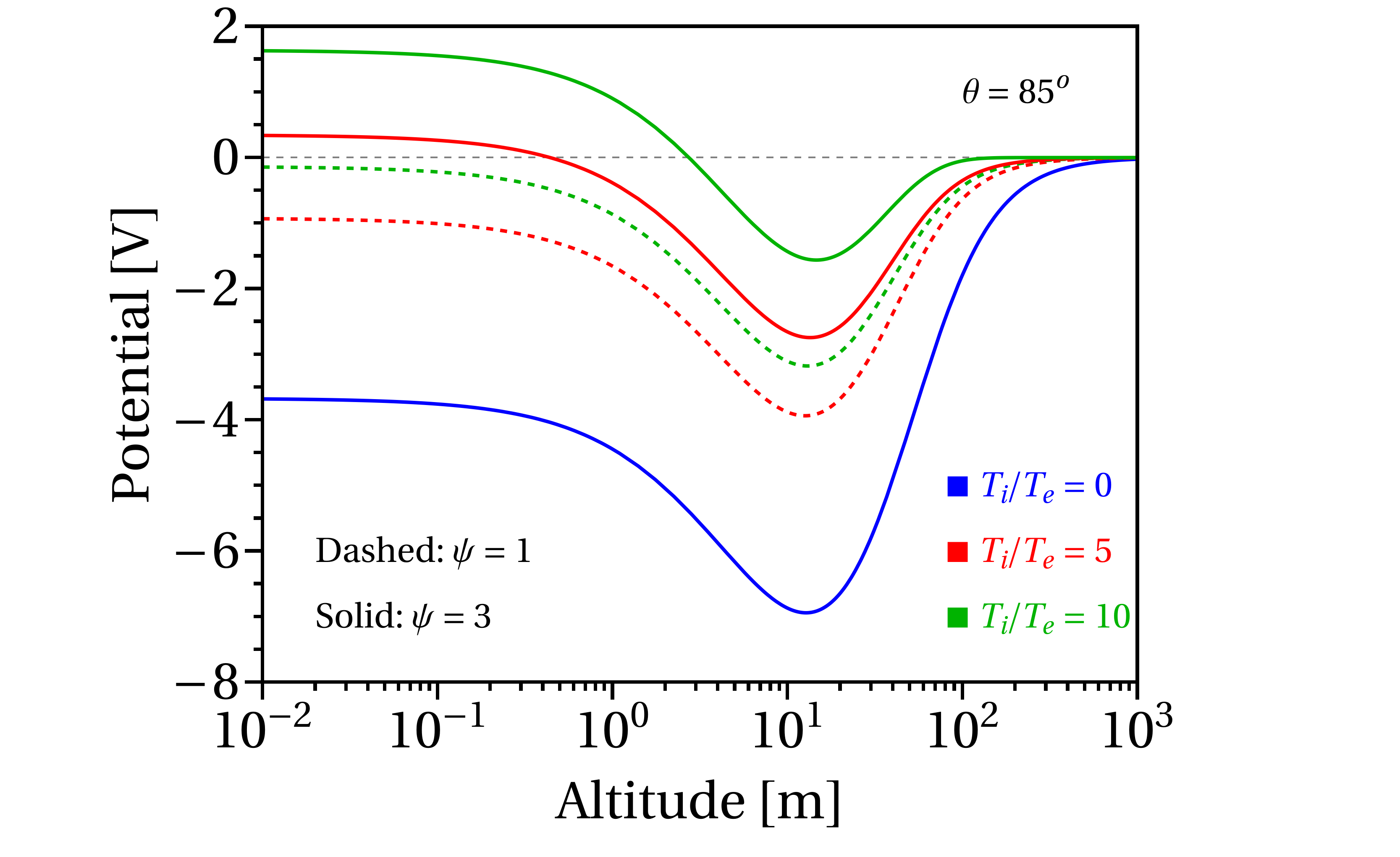}
    \caption{Altitude profile of non-monotonic potential structure at $\theta=85^\circ$ for $T_i/T_e=$ 0 (blue), 5 (red), and 10 (green). $\psi=1(3)$ for dashed (solid) curves. Other parameter used: $n_{i\infty}=5$ cm$^{-3}$, $T_e=15$ eV, and $J_{ph0}=9.7\,\mu $A m$^{-2}$.}
    \label{fig:Type A at 85 V}
\end{figure}

\begin{figure}
    \centering
    \includegraphics[width=1\linewidth]{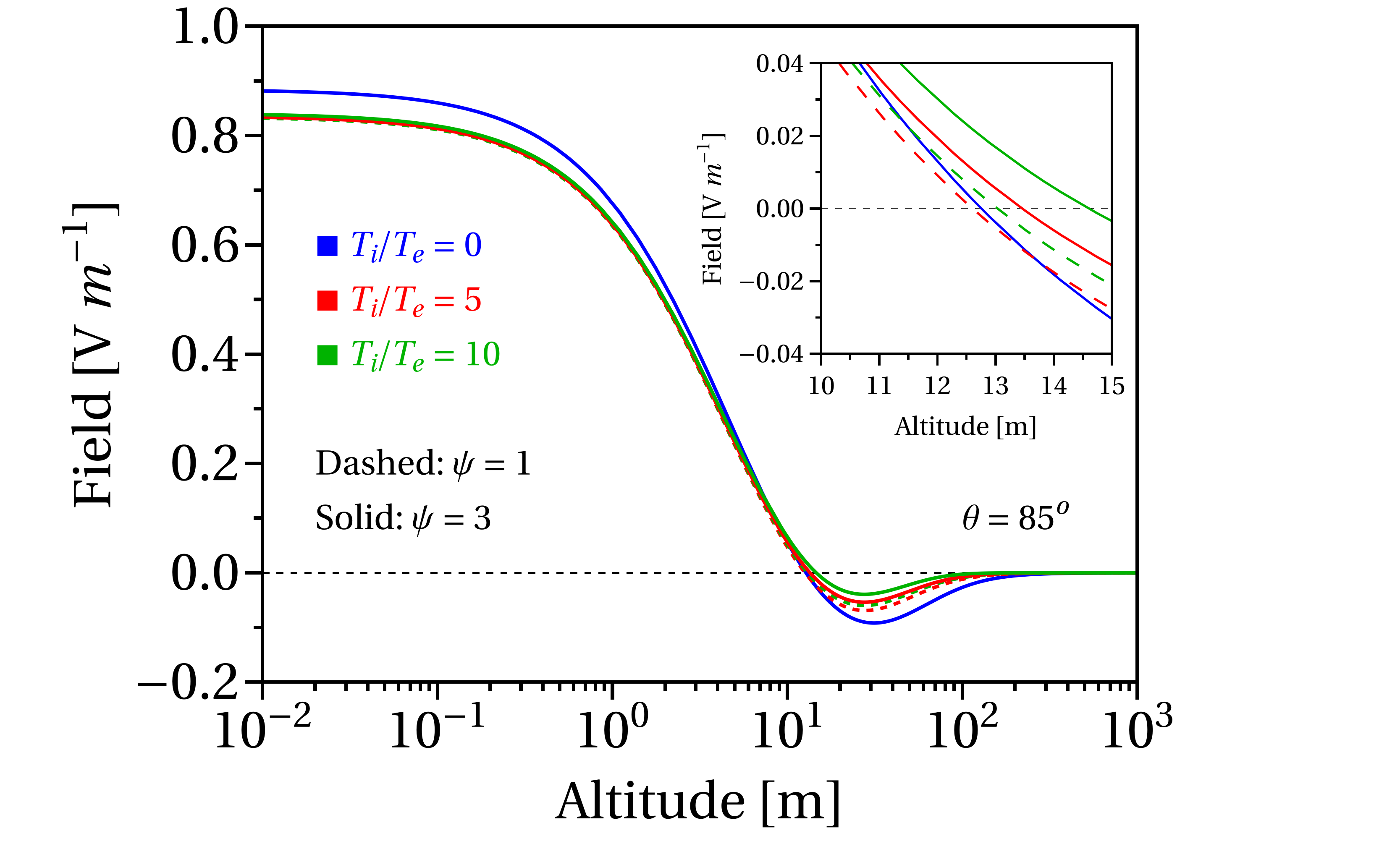}
    \caption{Altitude profile of non-monotonic field structure at $\theta=85^\circ$ for $T_i/T_e=$ 0 (blue), 5 (red), and 10 (green). $\psi=1(3)$ for dashed (solid) curves. Other parameter used: $n_{i\infty}=5$ cm$^{-3}$, $T_e=15$ eV, and $J_{ph0}=9.7\,\mu $A m$^{-2}$.}
    \label{fig:Type A at 85 E}
\end{figure}

\begin{figure}
    \centering
    \includegraphics[width=1\linewidth]{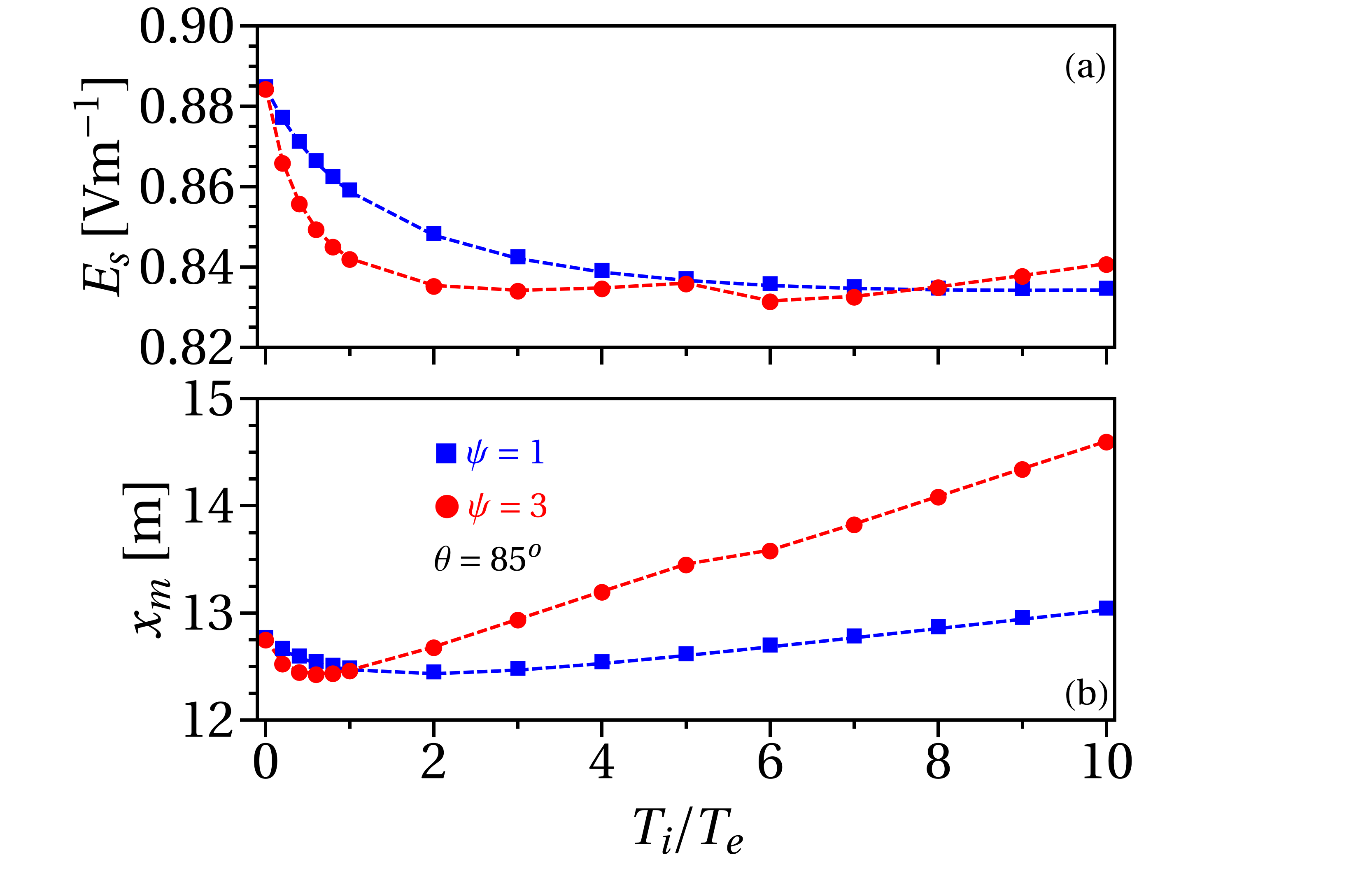}
    \caption{Variation of lunar surface electric field $(E_s)$ and location of potential minima $(x_m)$ with $T_i/T_e$ for non-monotonic potential structure at $\theta=85^\circ$. Other parameter used: $n_{i\infty}=5$ cm$^{-3}$, $T_e=15$ eV, and $J_{ph0}=9.7\,\mu $A m$^{-2}$.}
    \label{fig:xm and Es variation for Type A&B}
\end{figure}

\begin{figure}
    \centering
    \includegraphics[width=1\linewidth]{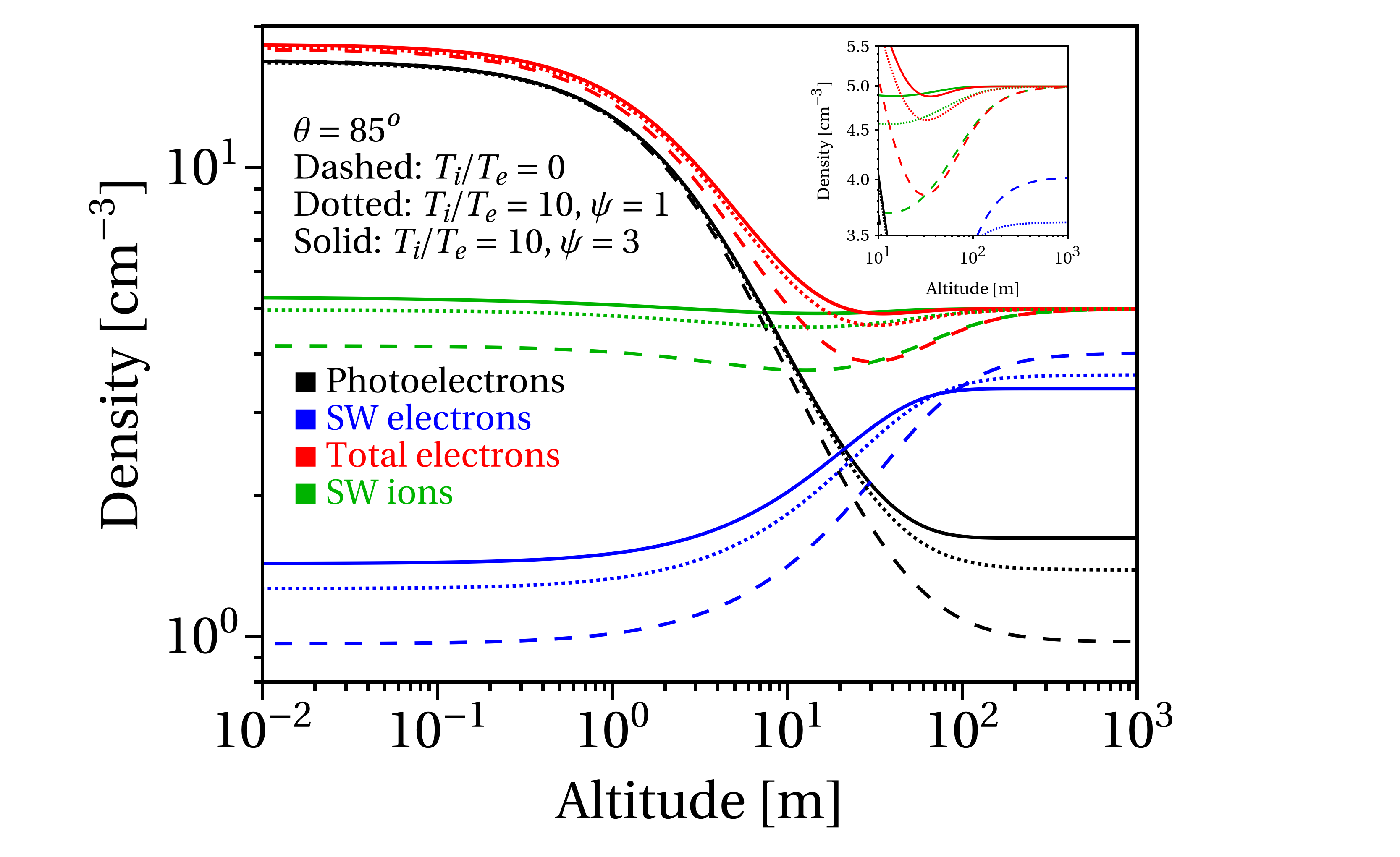}
    \caption{Altitude profile of population densities within non-monotonic sheath structure at $\theta=85^\circ$ for $T_i/T_e=$ and $\psi=1(3)$. Other parameter used: $n_{i\infty}=5$ cm$^{-3}$, $T_e=15$ eV, and $J_{ph0}=9.7\,\mu $A m$^{-2}$.}
    \label{fig:Type A at 85 den}
\end{figure}
The effect of $T_i$ on the non-monotonic potential and field structure at $\theta=85^\circ$ is illustrated in Figures \ref{fig:Type A at 85 V} and \ref{fig:Type A at 85 E} for three different $T_i/T_e$ values. A significant difference in the altitude profiles of $V(x)$ is observed. The location of $V_m$, i.e., $x_m$, is observed to be at different altitudes compared to $T_i/T_e=0$. Overall, the sheath thickness decreases with increasing $T_i/T_e$ and $\psi$ due to faster screening of the lower-magnitude $V_m$. However, below $x_m$, the gradient of potential, i.e., electric field, as illustrated in Figure \ref{fig:Type A at 85 E}, slightly reduces with $T_i/T_e$. The variation of surface electric field $(E_s)$ is illustrated in Figure \ref{fig:xm and Es variation for Type A&B}a. The reduction of $E_s$ is observed only in the second decimal places with increasing $T_i$. Although the trapped photoelectrons are mainly responsible for the electric field determination below $x_m$, an increase in $n_i$ near the surface is responsible for this slight field strength reduction. Furthermore, a notable variation in $x_m$ is observed in Figure \ref{fig:xm and Es variation for Type A&B}b, leading to a significant shift in the location of the zero electric field on a meter scale. Figure \ref{fig:Type A at 85 den} shows that the densities of warm ions within the sheath increase with increasing $T_i$ and $\psi$; and $n_i$ dominates within the sheath for $x>x_m$. This increase in ion density weakens $V_m$ and reduces $V_0-V_m$. As a result, photoelectron trapping and SW electron reflection within the sheath are reduced. This leads to a higher density of photoelectrons and SW electrons at the sheath edge and near the surface, respectively. Therefore, the significant change in sheath characteristics due to finite $T_i$ could influence the dust charging current within the sheath. Additionally, the variation in $x_m$ where the electric field reverses sign from positive to negative could significantly influence the dynamics of charged dust above the lunar surface. Overall, this study enhances our understanding of the lunar photoelectron sheath. It emphasizes that although ions contribute marginally to the lunar surface charging, they are essential for shaping the near-surface potential structure. Additionally, the dynamic variation in ion temperature observed by ARTEMIS could significantly affect the sheath structures and their subsequent detection.
%
%
\section{Summary}
The paper summarizes the results of numerical investigations of the lunar photoelectron sheath composed of photoelectrons, SW electrons, and ions. The photoelectrons and SW electrons are considered to be Maxwellian. A quantitative comparison of the sheath characteristics has been carried out for cold and warm ions. In the region with marginal photoemission (near the terminator and at higher latitudes), a classical Debye-type sheath structure develops above the lunar surface. On the other hand, near the equatorial and mid-latitudes, due to significant photoemission, a non-monotonic potential structure with a potential minimum is anticipated. The presence of a finite ion temperature significantly influences the quantitative estimates of these sheath structures. Notably, for the classical Debye-type potential structure in the lunar terminator region, finite $T_i$  
\begin{enumerate}
    \item decreases the magnitude of the negative potential.
    \item Reduces the sheath thickness.
    \item Reduces overall field strength within the sheath.
\end{enumerate}
For a non-monotonic potential structure with finite $T_i$  
\begin{enumerate}
 \setcounter{enumi}{3}
    \item significantly affects the potential structure at higher latitudes, where SW ions are entering the sheath with Bohm speed.
    \item Increases the lunar surface potential and reduces the strength of the negative potential minima.
    \item Increases the population densities of lunar photoelectrons and SW electrons near the sheath edge and in the vicinity of the surface, respectively. 
    \item Significantly affects the location of the potential minima over meter scales.
\end{enumerate}
Therefore, this analysis brings out the essence of finite ion temperature in modifying the sheath populations, charging currents, electric potential, and field within the photoelectron sheath over the sunlit lunar surface. It depicts that although SW ions have a marginal contribution to the lunar surface charging compared to photoemission and SW electron collection, they can significantly affect the overall structure of the lunar photoelectron sheath, particularly at higher latitudes. As a result, warm ions can significantly influence the estimates of dust charging and dynamics above the lunar surface. Furthermore, the predictions may be of practical significance in the detection of non-monotonic potential structure around the Moon from Lunar Prospector and THEMIS-ARTEMIS data\citep{poppe2011negative,poppe2012comparison}.
%
%
\appendix
%
%
\section{Expressions of $n_i$}\label{sec:Appendix nsi}
The $n_i$ can be derived from the following continuity and force equations in one dimension

\begin{eqnarray}
    \frac{\partial n_i}{\partial t}+\frac{\partial}{\partial x}\big(n_iu_i\big)&=&0\,,\label{eq:conti}\\
    m_in_i\left[\frac{\partial u_i}{\partial t}+u_i\frac{\partial u_i}{\partial x}\right]&=&-en_i\frac{\partial V}{\partial x}-\frac{\partial p_i}{\partial x}\,,\label{eq:Force}
\end{eqnarray}
where $m_i$, $u_i$, and $p_i$ are the ion mass, speed, and pressure, respectively. For isothermal flow, 

\begin{equation}
    p_i=n_ik_BT_i\,,\label{eq:p_i isothermal}
\end{equation}
where $T_i$ is the ion temperature. For adiabatic flow, the pressure term can be related to isothermal pressure as \citep{kumar2025simple} 

\begin{equation}
    \left(\frac{\partial p_i}{\partial n_i}\right)_S=\psi\left(\frac{\partial p_i}{\partial n_i}\right)_T\label{eq:isphermal to adiabatic}
\end{equation}
where the subscript $S(T)$ indicates constant entropy (temperature) of the system, and $\psi$ is the ratio of specific heats with constant pressure and volume called the adiabatic index. Based on the above equation for  adiabatic ion flow, we can write

\begin{equation}
    \frac{\partial p_i}{\partial x}=\psi k_BT_i\frac{\partial n_i}{\partial x}\,.\label{eq:dp/dx}
\end{equation}
Using the above, in steady state, Eq \ref{eq:Force} can be written as

\begin{equation}
    m_iu_i\frac{\partial u_i}{\partial x}=-e\frac{\partial V}{\partial x}-\frac{\psi k_BT_i}{n_i}\frac{\partial n_i}{\partial x}\,.\label{eq:steady state force}
\end{equation}
Integrating both sides from sheath edge $(x\rightarrow\infty)$ to finite $x$, we get

\begin{equation}
    \frac{1}{2} m_i\big(u_i^2-u_0^2\big)+\psi k_BT_i\ln\left(\frac{n_i}{n_{i\infty}}\right)+eV=0\,,\label{eq:energy conservation}
\end{equation}
where $u_0$ is the ion speed at the sheath edge. In steady state, the continuity equation shown in Eq. \ref{eq:conti} can be written as 

\begin{equation}
        \frac{\partial}{\partial x}\big(n_iu_i\big)=0\,,\label{eq:conti_steady}
\end{equation}
or,

\begin{equation}
    n_iu_i=n_{i\infty}u_0\,.\label{eq:Ion flux balance}
\end{equation}
Combining Eqs. \ref{eq:energy conservation}, \ref{eq:conti_steady} and \ref{eq:Ion flux balance} we can write

\begin{equation}
    n_i=n_{i\infty}\left[1-\frac{2k_BT_e}{m_iu_0^2}\left\{\frac{eV}{k_BT_e}+\psi\frac{T_i}{T_e}\ln\left(\frac{n_i}{n_{i\infty}}\right)\right\}\right]^{-1/2}\,.
\end{equation}
The above equation can be solved numerically for different values of $V$. 
%
%
\section{Expressions of $n_p$ and $n_e$}\label{sec:Appendix npe and nse}
Depending on the sheath profile, the population density corresponding to photoelectrons and  SW electrons within the sheath can be given by

\begin{equation}
    n_{j}\big(x\big)=\left[\int_{u_m(x)}^\infty+2\delta_j^c\int_{0}^{u_m(x)}\right]\int_{-\infty}^\infty\int_{-\infty}^\infty f_{j}(x,\,\mathbf{u})\, d^3\mathbf{u}
    \,,\label{eq:nsef}
\end{equation}
where ${u}_{m}(x)=\big(2e\big(V(x)-V_m\big)/m_e\big)^{1/2}$ is the minimum speed of the electrons at $x$ to overcome the potential minimum $V_m$ at $x_m$. Here, $j=p(e)$ represents photo(SW) -electrons. $m_e$ is the electron mass. For a non-monotonic potential structure $\delta_p^c=1$ for $x\le x_m$ and $\delta_p^c=0$ for $x> x_m$. And  $\delta_e^c=0$ for $x\le x_m$ and $\delta_e^c=1$ for $x> x_m$. For the classical Debye-type sheath structure, $x_m=0$ and $V_0=V_m$. Please see Ref. \cite{nitter1998} for more details. 

The Maxwellian distribution of photoelectrons and SW electrons at $x$ from the lunar surface can be expressed as\citep{nitter1998} 

\begin{equation}
    \begin{split}
        f_{j}(x,\,\mathbf{u})=&n_{j,\,\alpha_j}\left[\frac{m_e}{2\pi k_BT_j}\right]^{3/2}\times\\
        &\exp\left[-\frac{m_e\mathbf{u}^2}{2k_BT_j}+\frac{e\big(V(x)-V_{\alpha_j}\big)}{k_BT_j}\right]\,,
    \end{split}
\end{equation}
where $k_B$ is the Boltzmann constant. For $j=p$, $\alpha_p=0$, $V_0$ is the lunar surface potential (at $x=0$). $T_p$ is the photoelectron temperature. $n_{p0}/2$ number density of photoelectron going away from the surface at $x=0$. For $j=e$, $\alpha_e=\infty$, $V_\infty=0$ is the potential at the sheath edge ($x\rightarrow0$). $T_e$ is the SW electron temperature. $n_{e\infty}/2$ number density of SW electron going towards the surface from the sheath edge at $x\rightarrow0$. $n_{p0}$ can be derived from the photoemission current as

\begin{equation}
    n_{p0}\left[\frac{k_BT_p}{2\pi m_e}\right]^{1/2}=\frac{J_{ph0}}{e}\cos\theta\,,
\end{equation}
 where $J_{ph0}$ is the photoemission current from the uncharged lunar surface at $\theta=0^\circ$ (equator).
%
%
\section{Determination of $V_0$, $V_m$, $n_{e\infty}$, and $\eta$}\label{sec:BC}
$V_0$, $V_m$, and $n_{e\infty}$ have been derived based on the following considerations 

\begin{enumerate}
    \item We consider that charge neutrality is maintained at the sheath edge. i.e., at $x\rightarrow\infty$, $V\rightarrow 0$, the quasi-neutrality is maintained by the SW electrons/ions and photoelectrons, i.e.,

\begin{equation}
    n_p(\infty)+n_e(\infty)-n_i(\infty)=0\,.\label{eq:neutrality}
\end{equation}

\item We consider the sheath to be in a steady state with zero net current. The photoemission and SW ion collection charge the surface positively, while the SW electrons result in negative charging. In the steady state, these charging currents should be balanced within the sheath. Mathematically, it can be presented as

\begin{equation}
    I_p-I_e+I_i=0\,.
\end{equation}
At the sheath edge, the above equation becomes

\begin{equation}
\begin{split}
\int_{u_m(0)}^\infty&\int_{-\infty}^\infty\int_{-\infty}^\infty u_xf_p(0,\mathbf{u})\, d^3\mathbf{u}\\&+\int_{-\infty}^{-u_m(\infty)}\int_{-\infty}^\infty\int_{-\infty}^\infty u_xf_e(\infty,\mathbf{u})\, d^3\mathbf{u}\\ &\quad\quad+n_{i\infty}\max\left[u_B,u_{i\infty}\cos\theta\right]=0\,.
\end{split}
\label{eq:flux ballance}
\end{equation}
\item For a non-monotonic potential structure, a potential minimum exists at $x=x_m$. At this point, $dV/dx=0$. Multiplying both sides of Eq. \ref{eq:Basic Poisson Eq 1} by $2dV/dx$ and
using $dV/dx=0$ and $V=0$ as $x\rightarrow\infty$, one obtains another boundary condition specifically for a non-monotonic sheath structure as

\begin{equation}
    \int_{V_m}^0\big[n_i-n_e-n_p\big]dV=0\,.\label{eq:zero field}
\end{equation}
So, for non-monotonic sheaths, additional conditions have to be imposed at $x=x_m$: $V(x_m)=V_m$ and $-V'(x_m)=0$.
\end{enumerate}
For the Debye-type potential structure, $V_0$ and $n_{e\infty}$ can be derived from the simultaneous solution of Eqs. \ref{eq:neutrality} and \ref{eq:flux ballance}. 

For non-monotonic potential structures with $u_{i\infty}\cos\theta>u_B$, simultaneous solution of Eqs. \ref{eq:neutrality}, \ref{eq:flux ballance}, \ref{eq:zero field} provides the numerical estimates of $V_0$, $V_m$, and $n_{e\infty}$. In the case of $u_{i\infty}\cos\theta<u_B$, $u_B$ should be replaced by $\tilde u_B$ (given in Eq. \ref{eq:Bohm with push eta}). An additional equation has to be solved:

\begin{equation}
    F_{min}(\tilde u_B,\,V_0)=0\,.\label{eq:Fmin_solve_to_get_eta}
\end{equation}
The mathematical form of $F_{min}$ is given in Eq. \ref{eq:Fmin}. Eq. \ref{eq:Fmin_solve_to_get_eta} has to be solved to get the real values of $n_i(x)$ within the sheath, in the vicinity of the sunlit lunar surface. The above equation is essential for maintaining the ion continuity within the steady-state sheath structure. Thereafter, the simultaneous solution of Eqs. \ref{eq:neutrality}, \ref{eq:flux ballance}, \ref{eq:zero field}, and \ref{eq:Fmin_solve_to_get_eta} provides the numerical estimates $V_0$, $V_m$, $n_{e\infty}$, and $\eta$. 

Using these values, the potential structure can be derived by solving Poisson's equation given in Eq. \ref{eq:Basic Poisson Eq 1} following the computation schemes of Refs\cite{nitter1998,sana2023plasma}.
%
%
\section{Parameter used for calculation}\label{sec:parameter}
\begin{figure}
    \centering
    \includegraphics[width=1\linewidth]{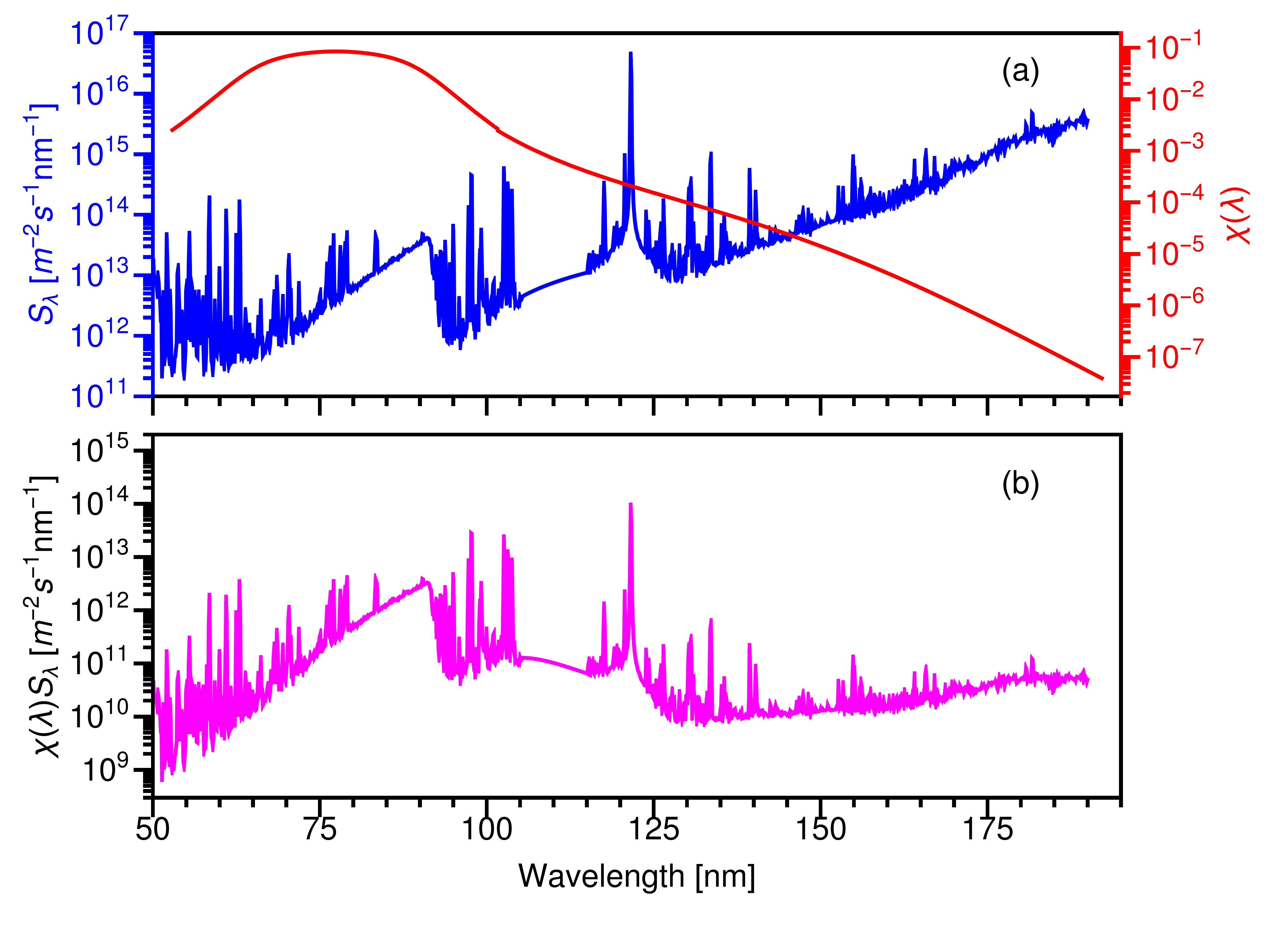}
    \caption{(a) Blue: Typical EUV emission spectrum $S_\lambda$ with at 1 AU from the Flare Irradiance Spectral Model 2 (FISM2)\citep{Chamberlin}. Red: The photoelectric yield $\chi(\lambda)$ measured by Ref. \cite{willis1973photoemission}. (b) Spectral dependence of the function $\chi(\lambda)S_\lambda$ representing the effective solar spectrum causing photoemission.}
    \label{fig:FISM2}
\end{figure}
\subsection*{$J_{ph0}$ and $T_p$}
The photoemission current from the uncharged lunar surface at $\theta=0^\circ$

\begin{equation}
    J_{ph0}=\int_{\lambda_\text{min}}^{\lambda_\text{max}}\chi(\lambda)S_\lambda d\lambda\,,
\end{equation}
where $S_\lambda$ is the typical UV/EUV emission spectrum, which is taken from the Flare Irradiance Spectral Model 2 (FISM2) measured at 1 AU\citep{Chamberlin} (shown in blue curve in Figure \ref{fig:FISM2}a). $\lambda_\text{min}$ is the minimum wavelength available in FISM2 spectrum. $\lambda_\text{max}$ is the maximum wavelength corresponding to the regolith work function. $\chi(\lambda)$ is the photoelectric yield of the lunar surface, which is taken from Ref.\cite{willis1973photoemission} following the measurements of Apollo lunar samples (shown in red curve in Figure \ref{fig:FISM2}a). The effective solar spectra giving rise to photoemission is shown in Figure \ref{fig:FISM2}b.
The work function of the lunar sample is taken to be 5 eV\cite{feuerbacher1972photoemission,willis1973photoemission}. For the spectra shown in Figure \ref{fig:FISM2}a, $ J_{ph0}=9.7\,\mu$A m$^{-2}$. Following Ref.\cite{sana2024velocity}, $T_p=3$ eV is taken in the present analysis.
\subsection*{SW plasma parameters}
Typical SW plasma parameter is taken to be $n_{i\infty}=5$ cm$^{-3}$, $T_e=15$ eV, and $u_{i\infty}=400$ km s$^{-1}$ at the sheath edge. And $T_i/T_e$ varied from 0 to 10 to highlight the effect of ion temperature.
%
%
\section{Determination of $V_c$ for warm ions}\label{sec:Vc for warm ions}
The determination of $V_c$ beyond which $n_i$ does not have a real value can be determined by combining the ion continuity and force equations. In a steady state, the ion speed $u_i$ at $x$ with potential $V$ can be derived from the equation

\begin{equation}
    \frac{1}{2} m_i\big(u_i^2-u_0^2\big)+\psi k_BT_i\ln\left(\frac{u_0}{u_i}\right)+eV=0\,,\label{eq:energy conservation_ui_root}
\end{equation}
or we can rewrite the above equation as

\begin{equation}
   F(u_i)=0\,.
\end{equation}
See Appendix \ref{sec:Appendix nsi} for the detailed derivation of the above equation. For fixed $u_0$, $\psi$, $T_i$, and $V$, from the real root of $F(u_i)$ (i.e., real value of $u_i$) $n_i$ can be numerically determined. Let's discuss the behavior of the function $F(u_i)$. Notice that $F(u_i)\rightarrow+\infty$ for $u_i\rightarrow0$ and $u_i\rightarrow\infty$. This suggests that $F(u_i)$ exhibits at least one minimum. Differentiating $F(u_i)$ with respect to $u_i$ and equating to zero

\begin{equation}
   \left. \frac{dF(u_i)}{du_i}\right|_{u_i=u_{im}}=0\,,
\end{equation}
or,

\begin{equation}
    u_{im}=\sqrt{\frac{\psi k_BT_i}{m_i}}\,.\label{eq:umin}
\end{equation}
Eq. \ref{eq:umin} suggests the location of the minima, because at $u_i=u_{im}$, we can easily show

\begin{equation}
     \left. \frac{d^2F(u_i)}{du_i^2}\right|_{u_i=u_{im}}>0\,.
\end{equation}
Hence,

\begin{equation}
    F_{min}= \frac{1}{2} m_i\left(\frac{\psi k_BT_i}{m_i}-u_0^2\right)+\psi k_BT_i\ln\left(\frac{u_0}{\sqrt{\frac{\psi k_BT_i}{m_i}}}\right)+eV\,.\label{eq:Fmin}
\end{equation}
$F(u_i)$ has real roots if and only if 

\begin{equation}
    F_{min}\le0\,.
\end{equation}
Otherwise, the real values of $u_i$ and $n_i$ do not exist at $x$ with potential $V(x)>0$. For supersonic values of $u_0$, $F_{min}<0$, and $F(u_i)$ exhibits two real roots for $V>0$. The root $u_i<u_0$ is the physically acceptable root, because the ion speed reduces near positive potential. Note that generalized $u_B$ given Eq. \ref{eq:Gen_Bohm_Speed} depends on $V_0$ and $V_m$. To have an approximate estimate of $V_c$, we are using the reference Bohm speed $u_{Br}=\sqrt{(k_BT_e+\psi k_BT_i)/m_i}$. Putting this in Eq. \ref{eq:Fmin}, we get

\begin{equation}
    V_c=\frac{k_BT_e}{2e}\left[1-\frac{\psi T_i}{T_e}\ln\left(1+\frac{T_e}{\psi T_i}\right)\right]\,.\label{eq:Vc_with_Ti}
\end{equation}
At the limit $T_i\rightarrow0$, the above equation reduces to Eq. \ref{eq:Vc_Cold_Ion} for cold ions.

\begin{figure}
    \centering
    \includegraphics[width=1\linewidth]{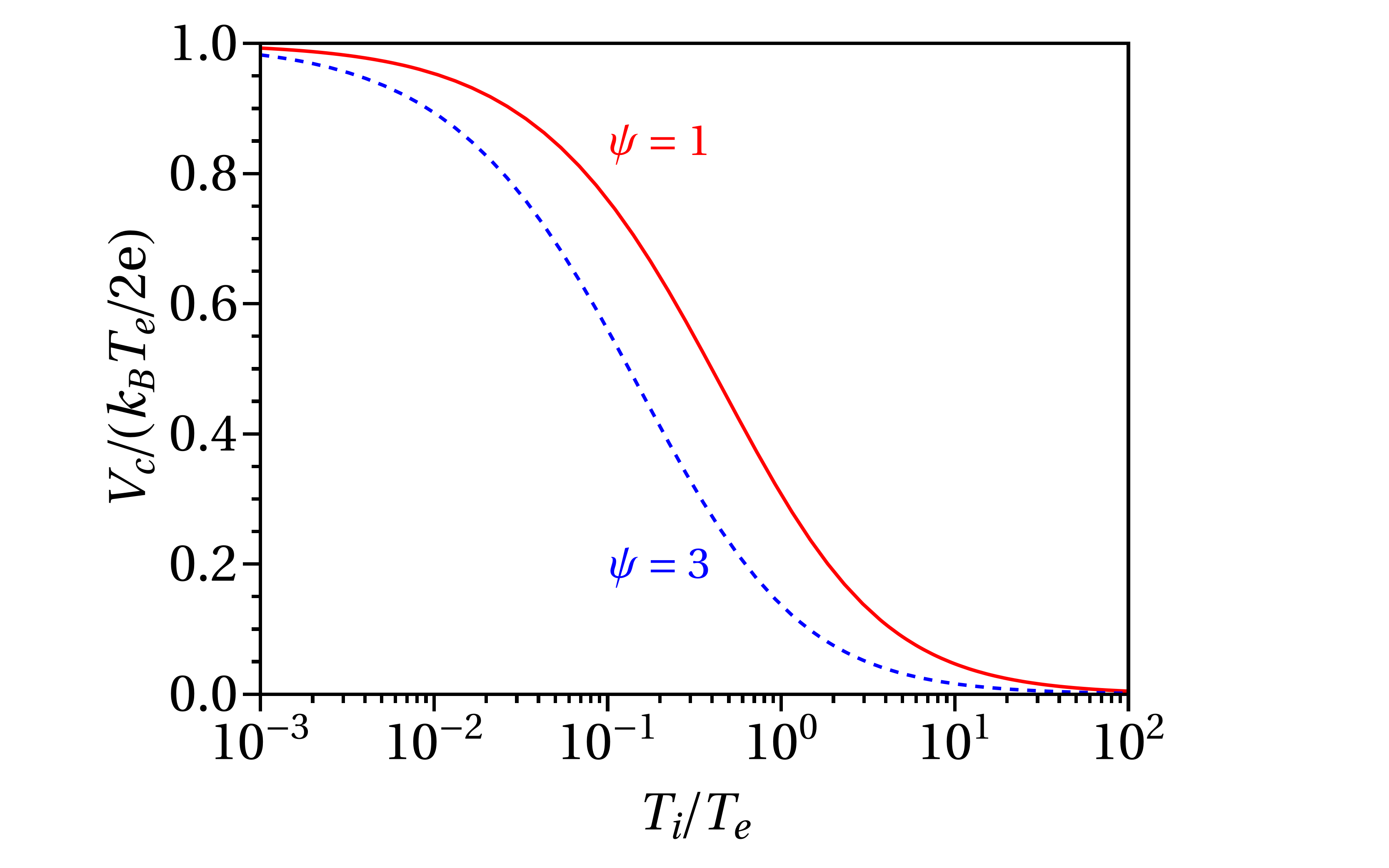}
    \caption{Variation of $V_c$ (normalized with $k_BT_e/2e$) with $T_i/T_e$.}
    \label{fig:Variation of Cut off Vc}
\end{figure}

\begin{figure}
    \centering
    \includegraphics[width=1\linewidth]{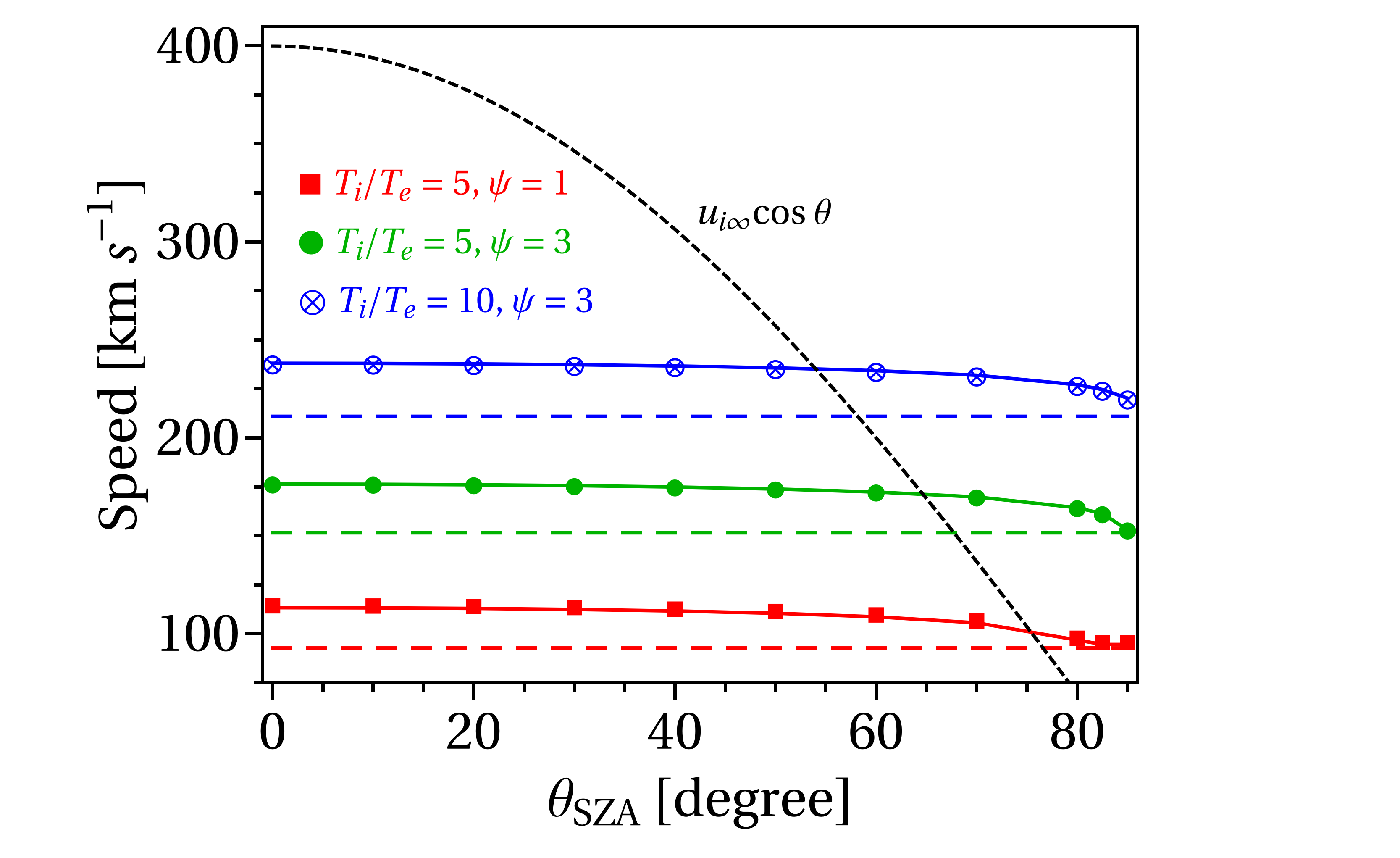}
    \caption{Variation of new Bohm speed $\tilde{u}_B$ (Solid) and reference Bohm speed $u_{Br}=\sqrt{(k_BT_e+\psi k_BT_i)/m_i}$ (Horizontal dashed lines), and SW speed normal to the surface (black curve) with $\theta$ for different $T_i/T_e$ and $\psi$.}
    \label{fig:Variation of Bohm speed.}
\end{figure}
Figure \ref{fig:Variation of Cut off Vc} illustrates that $V_c$ rapidly decreases with increasing $T_i/T_e$ and asymptotically reaches zero at very large $T_i/T_e$. 
\color{black}
It is important to note that $T_i$ results in an ion pressure gradient term. If the potential $V(x)$ at layer $x$ is positive relative to the sheath edge $(x\rightarrow\infty)$, then the ion speed at that layer is smaller than that of the sheath edge and $n_i(x)$ will be greater than $n_{i\infty}$ due to the continuity equation. In this layer, the ion thermal-pressure in the force equation acts toward the sheath edge. As ions move toward the lunar surface and pass through a region where $V(x) > 0$, they experience a push away from the surface and toward the sheath edge due to the thermal pressure and electrostatic repulsion. Consequently, as the thermal pressure increases with increasing $T_i$, even a small positive potential can create a significant force against the direction of ion motion. In that case, if the ions do not have sufficient drift speed, they cannot pass through that layer. As a result, $V_c$ decreases rapidly with increasing $T_i/T_e$.   
\color{black}
For the nominal SW condition with $T_e=15$ eV, the value of $V_c$ decreases from $7.5$ V for a cold ion to $0.4$ V (for $\psi=1$) and $0.1$ V (for $\psi=3$) for $T_i/T_e=10$, respectively. Due to the small positive value of $V_c$, $n_i$ may not have real values close to the surface in the region with significant photoemission if ions enter the sheath with $u_B$. This problem does not arise near the equatorial region because of the higher SW speed. However, at higher latitudes, the above problem may arise, which violates the ion continuity, and the sheath solution can not mathematically exist for $u_0=u_B$. Ions should enter the sheath edge with speed greater than or equal to $\tilde u_B$ to sustain the non-monotonic sheath structure. 

Figure \ref{fig:Variation of Bohm speed.} illustrates the modification of the Bohm speed at different $\theta$. In the parametric regime used in Figure \ref{fig:Variation of Bohm speed.}, for the equatorial and mid latitudes, $u_{i\infty}\cos\theta$ (the black curve) is well above the Bohm speed, leading to a stable non-monotonic sheath solution with finite $T_i$. However, at higher latitudes, ion speed falls below $\tilde u_B$ (see the black curve in Figure \ref{fig:Variation of Bohm speed.} falls below the Bohm speed curves). To form a stable sheath, ions must enter the sheath with $\tilde{u}_B$. It is further observed that $\tilde{u}_B$ does not differ significantly from $u_{Br}$, leading to small $\eta$ values. 
%
%
\section*{Acknowledgements}
This work was supported by the Department of Space, Government of India. We acknowledge NASA contract NAS5-02099 and V. Angelopoulos for the use of data from the THEMIS mission. TS is thankful to Dr. Prince Kumar for the helpful discussions. 
%
%
\section*{Data Availability}
The data from the THEMIS mission are available at \url{http://cdaweb.gsfc.nasa.gov/cdaweb/istp_public/}. The data from the FISM2 are available at \url{https://lasp.colorado.edu/lisird/data/fism_daily_hr}. The other data that support the findings of this study are available within the article.
%
%
\section*{References}
\bibliography{Ref}

@article{basnet2025generalized,
  title={Generalized Bohm sheath criterion for lunar dusty plasma and its interaction with lunar surface: Dust charging, dynamics, and levitation},
  author={Basnet, Suresh and Rai, Aman Kumar and Yoon, Young Dae and Misra, Amar Prasad},
  journal={Physics of Plasmas},
  volume={32},
  number={5},
  year={2025},
  publisher={AIP Publishing}
}

@article{riemann1991bohm,
  title={The Bohm criterion and sheath formation},
  author={Riemann, K-U},
  journal={Journal of Physics D: Applied Physics},
  volume={24},
  number={4},
  pages={493--518},
  year={1991}
}

@article{kumar2025simple,
  title={Simple fluid approach for the nonlinear excitations in Yukawa fluids},
  author={Kumar, Prince and Sharma, Devendra},
  journal={Journal of Plasma Physics},
  volume={91},
  number={2},
  pages={E61},
  year={2025},
  publisher={Cambridge University Press}
}

@article{poppe2012comparison,
  title={A comparison of ARTEMIS observations and particle-in-cell modeling of the lunar photoelectron sheath in the terrestrial magnetotail},
  author={Poppe, AR and Halekas, JS and Delory, GT and Farrell, WM and Angelopoulos, V and McFadden, JP and Bonnell, JW and Ergun, RE},
  journal={Geophysical research letters},
  volume={39},
  number={1},
  year={2012},
  publisher={Wiley Online Library}
}

@article{halekas2011first-non-mono,
  title={First remote measurements of lunar surface charging from ARTEMIS: Evidence for nonmonotonic sheath potentials above the dayside surface},
  author={Halekas, JS and Delory, GT and Farrell, WM and Angelopoulos, V and McFadden, JP and Bonnell, JW and Fillingim, MO and Plaschke, F},
  journal={Journal of Geophysical Research: Space Physics},
  volume={116},
  number={A7},
  year={2011},
  publisher={Wiley Online Library}
}

@article{poppe2011negative,
  title={Negative potentials above the day-side lunar surface in the terrestrial plasma sheet: Evidence of non-monotonic potentials},
  author={Poppe, Andrew and Halekas, Jasper S and Hor{\'a}nyi, Mih{\'a}ly},
  journal={Geophysical Research Letters},
  volume={38},
  number={2},
  year={2011},
  publisher={Wiley Online Library}
}

@article{singer1962photoelectric,
  title={Photoelectric screening of bodies in interplanetary space},
  author={Singer, SF and Walker, EH},
  journal={Icarus},
  volume={1},
  number={1-6},
  pages={7--12},
  year={1962},
  publisher={Elsevier}
}

@article{singer1962electrostatic,
  title={Electrostatic dust transport on the lunar surface},
  author={Singer, SF and Walker, EH},
  journal={Icarus},
  volume={1},
  number={1-6},
  pages={112--120},
  year={1962},
  publisher={Elsevier}
}

@article{sodhaandmishra2014,
  title={Lunar photoelectron sheath and levitation of dust},
  author={Sodha, MS and Mishra, SK},
  journal={Physics of Plasmas},
  volume={21},
  number={9},
  year={2014},
  publisher={AIP Publishing}
}

@article{nitter1998,
  title={Levitation and dynamics of charged dust in the photoelectron sheath above surfaces in space},
  author={Nitter, Tore and Havnes, Ove and Melands{\o}, Frank},
  journal={Journal of Geophysical Research: Space Physics},
  volume={103},
  number={A4},
  pages={6605--6620},
  year={1998},
  publisher={Wiley Online Library}
}

@article{walbridge1973lunar,
  title={Lunar photoelectron layer},
  author={Walbridge, Edward},
  journal={Journal of Geophysical Research},
  volume={78},
  number={19},
  pages={3668--3687},
  year={1973},
  publisher={Wiley Online Library}
}

@article{popel2013dusty,
  title={Dusty plasma at the surface of the Moon},
  author={Popel, SI and Kopnin, SI and Golub’, AP and Dol’nikov, GG and Zakharov, AV and Zelenyi, LM and Izvekova, Yu N},
  journal={Solar System Research},
  volume={47},
  pages={419--429},
  year={2013},
  publisher={Springer}
}

@article{popel2014distributions,
  title={On the distributions of photoelectrons over the illuminated part of the Moon},
  author={Popel, Sergei Igorevich and Golub, AP and Izvekova, Yu N and Afonin, VV and Dol’nikov, GG and Zakharov, AV and Zelenyi, Lev Matveevich and Lisin, EA and Petrov, Oleg Fedorovich},
  journal={JETP letters},
  volume={99},
  number={3},
  pages={115--120},
  year={2014},
  publisher={Springer}
}

@article{lisin2015lunar,
  title={Lunar dusty plasma: A result of interaction of the solar wind flux and ultraviolet radiation with the lunar surface},
  author={Lisin, EA and Tarakanov, VP and Popel, SI and Petrov, OF},
   journal={Journal of Physics: Conference Series},
  volume={653},
  number={1},
  pages={012139},
  year={2015},
  organization={IOP Publishing}
}

@article{mishra2019photoelectron,
  title={Photoelectron sheath on lunar sunlit regolith and dust levitation},
  author={Mishra, SK and Bhardwaj, A},
  journal={The Astrophysical Journal},
  volume={884},
  number={1},
  pages={5},
  year={2019},
  publisher={IOP Publishing}
}

@article{halekas2008,
  title={Lunar Prospector observations of the electrostatic potential of the lunar surface and its response to incident currents},
  author={Halekas, JS and Delory, GT and Lin, RP and Stubbs, TJ and Farrell, WM},
  journal={Journal of Geophysical Research: Space Physics},
  volume={113},
  number={A9},
  year={2008},
  publisher={Wiley Online Library}
}

@article{bohm1949,
  title={Minimum ionic kinetic energy for a stable sheath},
  author={Bohm, David},
  journal={The Characteristics of Electrical Discharges in Magnetic Fields},
  pages={77--86},
  year={1949},
  publisher={New York: McGraw-Hill}
}

@article{Chamberlin,
  title={The flare irradiance spectral model-version 2 ({FISM2})},
  author={Chamberlin, Phillip C and Eparvier, Francis G and Knoer, Victoria and Leise, H and Pankratz, Alicia and Snow, Martin and Templeman, Brian and Thiemann, Edward Michael Benjamin and Woodraska, Donald L and Woods, Thomas N},
  journal={Space Weather},
  volume={18},
  number={12},
  pages={e2020SW002588},
  year={2020},
  publisher={Wiley Online Library}
}

@article{mishra2020,
author = {Mishra,S. K. },
title = {Photoelectron distribution on sunlit surface of the Moon: A formalism},
journal = {Physics of Plasmas},
volume = {27},
number = {8},
pages = {082906},
year = {2020},
doi = {10.1063/5.0016411},

URL = { 
        https://doi.org/10.1063/5.0016411
    
},
eprint = { 
        https://doi.org/10.1063/5.0016411
    
}

}

@incollection{willis1973photoemission,
  title={Photoemission and secondary electron emission from lunar surface material},
  author={Willis, RF and Anderegg, M and Feuerbacher, B and Fitton, B},
  booktitle={Photon and particle interactions with surfaces in space},
  pages={389--401},
  year={1973},
  publisher={Springer}
}

@article{lisin2014effect,
  title={Effect of the solar wind on the formation of a photoinduced dusty plasma layer near the surface of the Moon},
  author={Lisin, EA and Tarakanov, VP and Petrov, OF and Popel, SI and Dol’nikov, GG and Zakharov, AV and Zelenyi, LM and Fortov, VE},
  journal={JETP letters},
  volume={98},
  number={11},
  pages={664--669},
  year={2014},
  publisher={Springer}
}

@article{dyadechkin2015new,
  title={New fully kinetic model for the study of electric potential, plasma, and dust above lunar landscapes},
  author={Dyadechkin, S and Kallio, E and Wurz, Peter},
  journal={Journal of Geophysical Research: Space Physics},
  volume={120},
  number={3},
  pages={1589--1606},
  year={2015},
  publisher={Wiley Online Library}
}

@article{poppe2010simulations,
  title={Simulations of the photoelectron sheath and dust levitation on the lunar surface},
  author={Poppe, Andrew and Hor{\'a}nyi, Mih{\'a}ly},
  journal={Journal of Geophysical Research: Space Physics},
  volume={115},
  number={A8},
  year={2010},
  publisher={Wiley Online Library}
}

@article{Fu_1970,
author = {Guernsey, Ralph L. and Fu, Jerry H. M.},
title = {Potential distribution surrounding a photo-emitting, plate in a dilute plasma},
journal = {Journal of Geophysical Research (1896-1977)},
volume = {75},
number = {16},
pages = {3193-3199},
doi = {https://doi.org/10.1029/JA075i016p03193},
url = {https://agupubs.onlinelibrary.wiley.com/doi/abs/10.1029/JA075i016p03193},
eprint = {https://agupubs.onlinelibrary.wiley.com/doi/pdf/10.1029/JA075i016p03193},
year = {1970}
}

@incollection{manka1973plasma,
  title={Plasma and potential at the lunar surface},
  author={Manka, Robert H},
  booktitle={Photon and particle interactions with surfaces in space},
  pages={347--361},
  year={1973},
  publisher={Springer}
}

@article{burinskaya2014influence,
  title={Influence of the solar wind on the distribution of the electric potential near the Moon’s surface},
  author={Burinskaya, TM},
  journal={Plasma Physics Reports},
  volume={40},
  number={1},
  pages={14--20},
  year={2014},
  publisher={Springer}
}

@article{burinskaya2015non,
  title={Non-monotonic potentials above the day-side lunar surface exposed to the solar radiation},
  author={Burinskaya, TM},
  journal={Planetary and Space Science},
  volume={115},
  pages={64--68},
  year={2015},
  publisher={Elsevier}
}

@article{zhao_photoelectron_2021,
	title = {Photoelectron {Sheath} and {Plasma} {Charging} on the {Lunar} {Surface}: {Semianalytic} {Solutions} and {Fully}-{Kinetic} {Particle}-in-{Cell} {Simulations}},
	volume = {49},
	issn = {1939-9375},
	shorttitle = {Photoelectron {Sheath} and {Plasma} {Charging} on the {Lunar} {Surface}},
	doi = {10.1109/TPS.2021.3110946},
	number = {10},
	journal = {IEEE Transactions on Plasma Science},
	author = {Zhao, Jianxun and Wei, Xinpeng and Du, Xiaoping and He, Xiaoming and Han, Daoru},
	month = oct,
	year = {2021},
	note = {Conference Name: IEEE Transactions on Plasma Science},
	pages = {3036--3050},
}

@inproceedings{feuerbacher1972photoemission,
  title={Photoemission from lunar surface fines and the lunar photoelectron sheath},
  author={Feuerbacher, B and Anderegg, M and Fitton, B and Laude, LD and Willis, RF and Grard, RJL},
  booktitle={Lunar and planetary science conference proceedings},
  volume={3},
  pages={2655},
  year={1972}
}

@article{stubbs2007lunar,
  title={Lunar surface charging: A global perspective using Lunar Prospector data},
  author={Stubbs, Timothy J and Halekas, Jasper S and Farrell, William M and Vondrak, Richard R},
  journal={Dust in planetary systems},
  volume={643},
  pages={181--184},
  year={2007},
  publisher={ESA Publications}
}

@article{sana2023plasma,
    author = {Sana, Trinesh and Mishra, S K},
    title = "{Plasma sheath around sunlit moon: monotonic and non-monotonic structures}",
    journal = {Monthly Notices of the Royal Astronomical Society},
    volume = {520},
    number = {1},
    pages = {233-246},
    year = {2023},
    month = {01},
    issn = {0035-8711},
}

@article{kumar2025aditya,
  title={Aditya solar wind particle experiment (ASPEX) on board Aditya—L1: the solar wind ion spectrometer (SWIS)},
  author={Kumar, Prashant and Bapat, Bhas and Shah, Manan S and Adalja, Hiteshkumar L and Patel, Arpit R and Adhyaru, Pranav R and Shanmugam, M and Chakrabarty, Dibyendu and Banerjee, Swaroop B and Subramanian, KP and others},
  journal={Solar Physics},
  volume={300},
  number={4},
  pages={37},
  year={2025},
  publisher={Springer}
}

@article{ogilvie1995swe,
  title={SWE, a comprehensive plasma instrument for the Wind spacecraft},
  author={Ogilvie, KW and Chornay, DJ and Fritzenreiter, RJ and Hunsaker, F and Keller, J and Lobell, J and Miller, G and Scudder, JD and Sittler Jr, EC and Torbert, RB and others},
  journal={Space Science Reviews},
  volume={71},
  number={1},
  pages={55--77},
  year={1995},
  publisher={Springer}
}

@article{angelopoulos_artemis_2011,
  title={The {ARTEMIS} {M}ission},
  author={Angelopoulos, Vassilis},
  journal={Space science reviews},
  volume={165},
  number={1},
  pages={3--25},
  year={2011},
  publisher={Springer}
}

@phdthesis{poppe2011modeling,
  title={Modeling, theoretical and observational studies of the lunar photoelectron sheath},
  author={Poppe, Andrew Reinhold},
  year={2011},
  school={University of Colorado at Boulder}
}

@article{fu_surface_1971,
  title={Surface potential of a photoemitting plate},
  author={Fu, Jerry HM},
  journal={Journal of Geophysical Research},
  volume={76},
  number={10},
  pages={2506--2509},
  year={1971},
  publisher={Wiley Online Library}
}

@article{nitter_dynamics_1992,
	title = {Dynamics of dust in a plasma sheath and injection of dust into the plasma sheath above Moon and asteroidal surfaces},
	volume = {56},
	number = {1},
	journal = {Earth, Moon and Planets},
	author = {Nitter, Tore and Havnes, Ove},
	month = jan,
	year = {1992},
	pages = {7--34},
}

@article{li2016dust,
  title={Dust levitation and transport over the surface of the Moon},
  author={Li, Lei and Zhang, YiTeng and Zhou, Bin and Feng, YongYong},
  journal={Science China Earth Sciences},
  volume={59},
  pages={2053--2061},
  year={2016},
  publisher={Springer}
}

@article{sana2024velocity,
  title={Velocity distribution of photoelectrons over sunlit moon},
  author={Sana, Trinesh and Mishra, SK},
  journal={Icarus},
  pages={115996},
  year={2024},
  publisher={Elsevier}
}

@article{grard1971photoelectron,
  title={Photoelectron sheath near a planar probe in interplanetary space},
  author={Grard, RJL and Tunaley, JKE},
  journal={Journal of Geophysical Research},
  volume={76},
  number={10},
  pages={2498--2505},
  year={1971},
  publisher={Wiley Online Library}
}

@article{wang2008modeling,
  title={Modeling electrostatic levitation of dust particles on lunar surface},
  author={Wang, Joseph and He, Xiaoming and Cao, Yong},
  journal={IEEE transactions on plasma science},
  volume={36},
  number={5},
  pages={2459--2466},
  year={2008},
  publisher={IEEE}
}

\end{document}